\documentclass[12pt,preprint]{aastex}

\usepackage{natbib}
\usepackage{graphicx,amsmath,amsfonts,amssymb}

\shorttitle{Towards $\sim1000$ Strongly Lensed Galaxies}
\shortauthors{Gonz\'alez-Nuevo et al.}

\begin{document}
\def\lsim{\mathrel{\rlap{\lower 3pt\hbox{$\sim$}}\raise 2.0pt\hbox{$<$}}}
\def\gsim{\mathrel{\rlap{\lower 3pt\hbox{$\sim$}} \raise 2.0pt\hbox{$>$}}}


\renewcommand{\thefootnote}{\fnsymbol{footnote}}

\title{\textit{Herschel}\footnote{{\it Herschel} is an ESA space observatory with science instruments provided by European-led Principal Investigator consortia and with important participation from NASA.}\,--ATLAS: towards a sample of $\sim1000$ strongly-lensed galaxies}


\author{J. Gonz\'alez-Nuevo\altaffilmark{1,19},
    A. Lapi\altaffilmark{2,1},
	S. Fleuren\altaffilmark{3},	
    S. Bressan\altaffilmark{4,1},
    L. Danese\altaffilmark{1},
	G. De Zotti\altaffilmark{4,1},
	M. Negrello\altaffilmark{5},
	Z.-Y. Cai\altaffilmark{1},
	L. Fan\altaffilmark{1},
	W. Sutherland\altaffilmark{3},
    M. Baes\altaffilmark{6},		
    A.J. Baker\altaffilmark{33},
    D.L. Clements\altaffilmark{26},
	A. Cooray\altaffilmark{8},
	H. Dannerbauer\altaffilmark{9},
    L. Dunne\altaffilmark{11},
    S. Dye\altaffilmark{11},	
    S. Eales\altaffilmark{12},
    D.T. Frayer\altaffilmark{31},
    A.I. Harris\altaffilmark{32},
    R. Ivison\altaffilmark{13,14},	
	M.J. Jarvis\altaffilmark{15,16},    				
	M.J. Micha{\l}owski\altaffilmark{14},
	M. L\'opez-Caniego\altaffilmark{19},
    G. Rodighiero\altaffilmark{29},		
	K. Rowlands\altaffilmark{11},
	S. Serjeant\altaffilmark{5},
    D. Scott\altaffilmark{30},
    P. van der Werf\altaffilmark{21},
	R. Auld\altaffilmark{12},
	S. Buttiglione\altaffilmark{4},
    A. Cava\altaffilmark{23},
    A. Dariush\altaffilmark{24,28},
    J. Fritz\altaffilmark{25},
    R. Hopwood\altaffilmark{5,26},
    E. Ibar\altaffilmark{13},
    S. Maddox\altaffilmark{11},
    E. Pascale\altaffilmark{12},
    M. Pohlen\altaffilmark{12},
    E. Rigby\altaffilmark{11},
    D. Smith\altaffilmark{15},
    P. Temi\altaffilmark{27}
    }
\email{gnuevo@ifca.unican.es}


\altaffiltext{1}{SISSA, Via Bonomea 265, I-34136 Trieste, Italy}
\altaffiltext{2}{Dipartimento di Fisica, Universit\`a di Roma `Tor Vergata', Via Ricerca Scientifica 1, 00133 Roma, Italy}
\altaffiltext{3}{School of Mathematical Sciences, Queen Mary, University of London, Mile End Road, London, E1 4NS, UK}
\altaffiltext{4}{INAF-Osservatorio Astronomico di Padova, Vicolo dell'Osservatorio 5, I-35122 Padova, Italy}
\altaffiltext{5}{Department of Physical Sciences, The Open University, Milton Keynes MK7 6AA, UK}
\altaffiltext{6}{Sterrenkundig Observatorium, Univ. Gent, Krijgslaan 281 S9,
B-9000 Gent, Belgium}
\altaffiltext{8}{Dept. of Physics \& Astronomy, Univ. of California, Irvine, CA 92697, USA}
\altaffiltext{9}{Universit\"at Wien, Institut f\"ur Astronomie, T\"urkenschanzstra{\ss}e 17, 1180 Wien, \"{O}sterreich}
\altaffiltext{11}{School of Physics and Astronomy, Univ. of Nottingham, Univ. Park, Nottingham NG7 2RD, UK}
\altaffiltext{12}{School of Physics and Astronomy, Cardiff Univ., The Parade, Cardiff, CF24 3AA, UK}
\altaffiltext{13}{UK Astronomy Technology Centre, Royal Observatory, Blackford Hill, Edinburgh EH9 3HJ, UK}
\altaffiltext{14}{Inst. for Astronomy, Univ. of Edinburgh, Royal Observatory, Blackford Hill, Edinburgh EH9 3HJ, UK}
\altaffiltext{15}{Centre for Astrophysics, Univ. of Hertfordshire, Hatfield, Herts, AL10 9AB, UK}
\altaffiltext{16}{Physics Dept., Univ. of the Western Cape, Cape Town, 7535, South Africa}

\altaffiltext{19}{Inst. de Fisica de Cantabria (CSIC-UC), Avda. los Castros s/n, 39005 Santander, Spain}
\altaffiltext{21}{Leiden Observatory, Leiden University, P.O. Box 9513, NL - 2300 RA Leiden, The Netherlands}
\altaffiltext{23}{Dep. de Astrof\'{\i}sica, Fac. de CC. F\'{\i}sicas, Univ. Complutense de Madrid, E-28040 Madrid, Spain}
\altaffiltext{24}{Physics Department, Imperial College London, South Kensington Campus, London SW7 2AZ, UK}
\altaffiltext{25}{Sterrenkundig Observatorium, Univ. Gent, Krijgslaan 281 S9, B-9000 Gent, Belgium}
\altaffiltext{26}{Astrophysics Group, Imperial College, Blackett Lab, Prince Consort Road, London SW7 2AZ, UK}
\altaffiltext{27}{Astrophysics Branch, NASA Ames Research Center, Mail Stop 245-6, Moffett Field, CA 94035, USA}
\altaffiltext{28}{School of Astronomy, Institute for Research in Fundamental Sciences (IPM), PO Box 19395-5746, Tehran, Iran}
\altaffiltext{29}{Dipartimento di Astronomia, Universit\`a di Padova, Vicolo dell'Osservatorio 3, I-35122 Padova, Italy}
\altaffiltext{30}{Department of Physics \& Astronomy, University of British Columbia, 6224 Agricultural Road, Vancouver, BC, V6T1Z1, Canada}
\altaffiltext{31}{National Radio Astronomy Observatory, P.O. Box 2, Green Bank, WV 24944, USA}
\altaffiltext{32}{Department of Astronomy, University of Maryland, College Park, MD 20742, USA}
\altaffiltext{33}{Department of Physics and Astronomy, Rutgers, the State University of New Jersey, 136 Frelinghuysen Road, Piscataway, NJ 08854-8019, USA}


\begin{abstract}
While the selection of strongly lensed galaxies with $500\mu$m flux density $S_{500}> 100\,$mJy has proven to be rather straightforward (Negrello et al. 2010), for many applications it is important to analyze samples larger than the ones obtained when confining ourselves to such a bright limit. Moreover, only by probing to fainter flux densities is possible to exploit strong lensing to investigate the bulk of the high-$z$ star-forming galaxy population. We describe HALOS (the \textit{Herschel}-ATLAS Lensed Objects Selection), a method for efficiently selecting fainter candidate strongly lensed galaxies, reaching a surface density of $\simeq 1.5$--$2\,\hbox{deg}^{-2}$, i.e. a factor of about 4 to 6 higher than that at the 100 mJy flux limit. HALOS will allow the selection of up to $\sim 1000$ candidate strongly lensed galaxies (with amplifications $\mu\gtrsim2$) over the full H-ATLAS survey area. Applying HALOS to the H-ATLAS Science Demonstration Phase (SDP) field ($\simeq 14.4\,\hbox{deg}^{2}$) we find 31 candidate strongly lensed galaxies, whose candidate lenses are identified in the VIKING near-infrared catalog. Using the available information on candidate sources and candidate lenses we tentatively estimate a $\simeq 72\%$ purity of the sample. As expected, the purity decreases with decreasing flux density of the sources and with increasing angular separation between candidate sources and lenses. The redshift distribution of the candidate lensed sources is close to that reported for most previous surveys for lensed galaxies, while that of candidate lenses extends to substantially higher redshifts than found in the other surveys. The counts of candidate strongly lensed galaxies are also in good agreement with model predictions (Lapi et al. 2011). Even though a key ingredient of the method is the deep near-infrared VIKING photometry, we show that H-ATLAS data alone allow the selection of a similarly deep sample of candidate strongly lensed galaxies with an efficiency close to 50\%; a slightly lower surface density ($\simeq 1.45\,\hbox{deg}^{-2}$) can be reached with a $\sim 70\%$ efficiency.
\end{abstract}





\section{Introduction}\label{sec:intro}

As stressed by \citet{Treu10} most of the applications of strong gravitational lensing to address major astrophysical and cosmological issues are currently limited by sample size. Samples of thousands of strongly lensed systems are needed to make substantial progress. This will indeed be a major task for future wide field optical \citep[see, e.g.,][]{Oguri10} and radio (SKA) surveys \citep[e.g.,][]{Koopmans04}.

However, as predicted by \citet{Blain96}, \citet{Perrotta02,Perrotta03}, \citet{Negrello07}, \citet{Paciga2009}, and \citet{Lima2010}, among others, and  demonstrated by \citet{Negrello10},  millimeter and sub-millimeter surveys are an especially effective route to reach this goal.

This is because the counts of unlensed high-$z$ \mbox{(sub-)mm} galaxies (SMGs) drop very rapidly at bright flux densities, mirroring the rapid build-up of proto-spheroidal galaxies \citep{Granato04,Lapi11}. The magnification bias of the counts due to gravitational lensing is then boosted, making the selection of strongly lensed galaxies particularly easy for relatively shallow large area (sub-)mm surveys.

For example, objects above $100\,$mJy at $500\,\mu$m were predicted \citep{Negrello07} to comprise almost equal numbers of low-$z$ ($z\le 0.1$) late-type galaxies, with far-IR emission well above the IRAS detection limit and easily identified in the optical, and high-$z$ ($z>1$) strongly lensed SMGs, plus a handful of radio sources (mostly blazars), also easily identified in low-frequency radio catalogs. The predicted (and observed) surface density of strongly lensed galaxies (SLGs) with $500\mu$m flux density brighter than $S_{500}=100\,$mJy is $\simeq 0.3\,\hbox{deg}^{-2}$. A similar surface density of candidate SLGs was found by \citet{Vieira10} at the detection limit of their  $87\,\hbox{deg}^2$ survey with the South Pole Telescope (SPT) at 1.4 and 2 mm. This means that the SPT, which plans to cover some 2,500$\,\hbox{deg}^{2}$, may yield a sample of $\simeq 750$ SLGs.

The \emph{Herschel} Astrophysical Terahertz Large Area Survey\footnote{http://www.h-atlas.org/} \citep[H-ATLAS;][]{Eales10}, the largest area survey carried out by the \emph{Herschel} Space Observatory  \citep{Pilbratt10}, covering $\sim 550\,\mathrm{deg}^2$ with PACS \citep{Poglitsch10} and SPIRE \citep{Griffin10}, will easily provide a sample of  about 150--200 SLGs with $S_{500}\ge 100\,$mJy. Many more such objects may be found at fainter flux densities, but singling them out is more difficult because they are mixed with high-$z$ unlensed galaxies.

The selection of fainter SLGs has the important additional bonus that it allows us to pick up galaxies more representative of the bulk of the star-forming galaxy population at $z \simeq $1--3. High-$z$ SLGs brighter than 100\,mJy at $500\,\mu$m have apparent far-IR luminosities ${\rm L}_{\rm FIR}> 3\times 10^{13}\,{\rm L}_\odot$ \citep{Negrello10}. Correcting for a gravitational amplification by a factor of 10  \citep[typical of these sources, see ][]{Harris12}, their far-IR luminosity corresponds to a star-formation rate ${\rm SFR}> 500\,{\rm M}_\odot\,\hbox{yr}^{-1}$. In contrast, data from sensitive near-infrared integral field spectrometers mounted on 8-10m class telescopes \citep[e.g. ][]{Forster09} suggest that the most effective star formers in the Universe have high but far less extreme SFRs ($\hbox{SFR}\sim 100\hbox{--}200\,{\rm M}_\odot\,\hbox{yr}^{-1}$). The power of strong lensing is needed to detect these sources, that are otherwise well below the SPIRE confusion limit; but we need to select SLGs with sub-mm flux densities as faint as possible.

In this paper we discuss a strategy that exploits the multi-wavelength coverage of the H-ATLAS survey areas to improve the selection efficiency of candidate SLGs fainter than $S_{500}=100\,$mJy. We apply our strategy to objects detected in the H-ATLAS Science Demonstration Phase (SDP) field, that covers an area of $\approx 3.8^\circ\times 3.8^\circ$ centered on $(\alpha,\delta)= $ $(09^{\rm h}\,05^{\rm m},$ $+0^\circ\, 30';$ J2000) to the same depth as the general H-ATLAS survey. Complete descriptions of the reduction of PACS and SPIRE SDP data are given in \citet{Ibar10} and \citet{Pascale11}, respectively. Source extraction and flux density estimation are described in \citet{Rigby11}. The $5\sigma$ detection limits,  including confusion noise, are $33.5$, 37.7, and $44.0\,\hbox{mJy}\,\hbox{beam}^{-1}$ in the SPIRE bands at 250, 350, and $500\,\mu$m, respectively; in the PACS bands they are $132\,\hbox{mJy}\,\hbox{beam}^{-1}$ and $121\,\hbox{mJy}\,\hbox{beam}^{-1}$ at 100 and $160\,\mu$m, respectively \citep{Rigby11}.

We show that this strategy can allow us to reach candidate SLGs surface densities of $\sim 1.5$--$2\,\hbox{deg}^{-2}$, that would imply a total of up to $\sim 1000$ SLGs in the full H-ATLAS survey. The outline of this paper is as follows. In \S\,\ref{sec:sample} we describe the selection of the parent sample, that, in \S\,\ref{sec:LF}, is exploited to re-assess the bright end of the luminosity function in the same redshift bins as in \citet{Lapi11}. In \S\,\ref{sec:optcrit} we describe our approach to single out candidate SLGs in the parent sample and to estimate the purity of the candidate SLG sample. While our method relies on the deep near-IR photometry provided by the VISTA Kilo-degree INfrared Galaxy survey \citep[VIKING; ][]{Sutherland11,Fleuren11}, in \S\,\ref{sec:lumsel} we show how a selection with only a modest efficiency loss can be achieved using H-ATLAS data alone. Our main results are summarized and discussed in \S\,\ref{sec:discussion}.

\section{Sample selection}\label{sec:sample}

\citet{Lapi11} have selected a sample of candidate high-redshift ($z\ge 1.2$) H-ATLAS SDP galaxies starting from a sample of objects obeying the following criteria: i) $S_{250}\ge 35\,$mJy; ii) no Sloan Digital Sky Survey (SDSS) counterpart with reliability $R>0.8$, as determined by \citet{Smith11}; and iii) $\ge 3\sigma$ detection at $350\,\mu$m. As pointed out in that paper, this sample is biased against strongly lensed galaxies that may have an apparently reliable SDSS identification (at a very small angular separation), which, however, is the foreground lens. This was a minor problem for the purpose of the Lapi et al. paper, but for the purpose of the present paper these objects need to be recovered. This can be done by checking whether the optical luminosities and colors of the possible identifications are compatible with the {\it Herschel} photometric data. The task is eased by the fact that frequently \citep[but not always, as demonstrated by the SWELLS survey,][]{Treu11} the lenses are passive elliptical galaxies \citep{Auger09,Negrello10}.

\begin{figure}
\begin{center}
\includegraphics[width=1\textwidth]{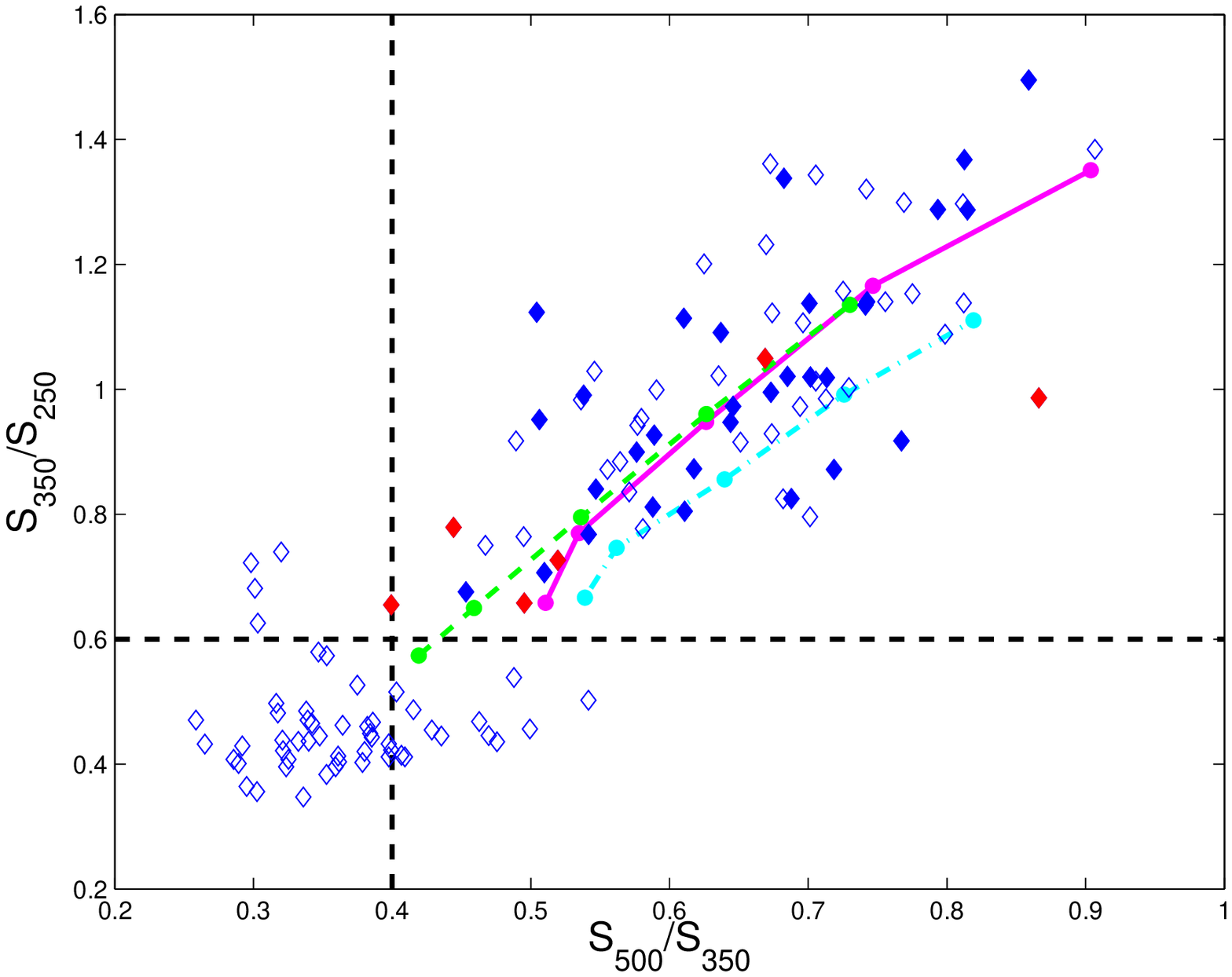}
\caption{SPIRE colors of objects with $S_{350}>85\,$mJy and $S_{250}>35\,$mJy (blue empty diamonds). The black dashed lines indicate the limits used for the selection of high redshift objects. The variation with redshift of the colors for the SEDs of three ultraluminous infrared galaxies (SMM J2135$-0102$, green dashed; Arp220, cyan dot-dashed; G15.141, magenta solid) are shown for comparison;  the filled circles along the lines correspond to $z=[1.2, 1.5, 2.0, 2.5, 3.0]$, with $z$ increasing from the lower left to the upper right corner. The blue filled diamonds are the strong SLG candidates identified in the SDP field (see \S\,\ref{sec:optcrit}). The red filled diamonds are objects dropped from the initial sample (see text for more details; one of the dropped objects is not shown because its colors are out of range).}
\label{fig:ircolor}%
\end{center}
\end{figure}

Using the formalism of \citet{Perrotta02} and the sub-mm luminosity functions of \citet{Lapi11} we find that a substantial increase in the surface density of strongly lensed sources can be achieved, still ensuring that the ratio of lensed to unlensed high-$z$ galaxies is not far below unity, by applying a flux density cutoff of 85 mJy at $350\,\mu$m. At this limit, the model yields surface densities of $\simeq 2\,\hbox{deg}^{-2}$, for both lensed and unlensed high-$z$ galaxies (while the surface density of $z\lsim 1$ galaxies is $\simeq 4\,\hbox{deg}^{-2}$). In the SDP field we have 127 objects with $S_{350}\ge 85\,$mJy and $S_{250}\ge 35\,$mJy. Their SPIRE colors are plotted in Fig.~\ref{fig:ircolor}, along with the colors yielded by the Spectral Energy Distributions (SEDs) of three ultraluminous dusty galaxies (Arp 220; SMM~J2135-0102, `The Cosmic Eyelash' at $z=2.3$, \citealt{Ivison10}, \citealt{Swinbank10}; H-ATLAS J142413.9+022304 alias G15.141 at $z=4.23$, \citealt{Cox11}) as a function of redshift. For all the three SEDs, objects with $S_{350}/S_{250}>0.6$ and $S_{500}/S_{350}>0.4$ are at $z_{\rm source}\ge 1.2$. Nevertheless, as discussed in \citet{Lapi11}, some of them might be low-$z$ galaxies with moderate SFRs and cold far-IR SEDs, but in that case they would be expected to have SDSS counterparts.

There are 74 objects, out of the total of 127, that conform to these color criteria. Two of them, however, have anomalous colors and were excluded from the subsequent analysis: HATLAS J090402.9+005436 (SDP.34) is a compact Galactic molecular cloud, also known as the ``H-ATLAS Blob'' \citep{Thompson11}; and HATLAS J090025.4-003019 (SDP.218) has $S_{350}/S_{250}<1.1\times S_{500}/S_{350}$, perhaps indicating a substantial boosting of the $500\,\mu$m flux density due to a background fluctuation.
We also excluded HATLAS J090910.1+012135 (SDP.61) because it is a blazar \citep[][it does not show up in Fig.~\ref{fig:ircolor} because its $S_{500}/S_{350}$ color is out of range]{gnuevo10}. We further exclude HATLAS J090923.9+000210 (SDP.362) because it is a QSO.
Three more objects (HATLAS J090359.6-004556 = SDP.70, HATLAS J085828.4+012210 = SDP.85, and HATLAS J091059.1+000303 = SDP.121) were also excluded because they have PACS flux densities that suggest $z<1$.

\begin{figure}
\begin{center}
  \centering
  \includegraphics[width=0.45\textwidth]{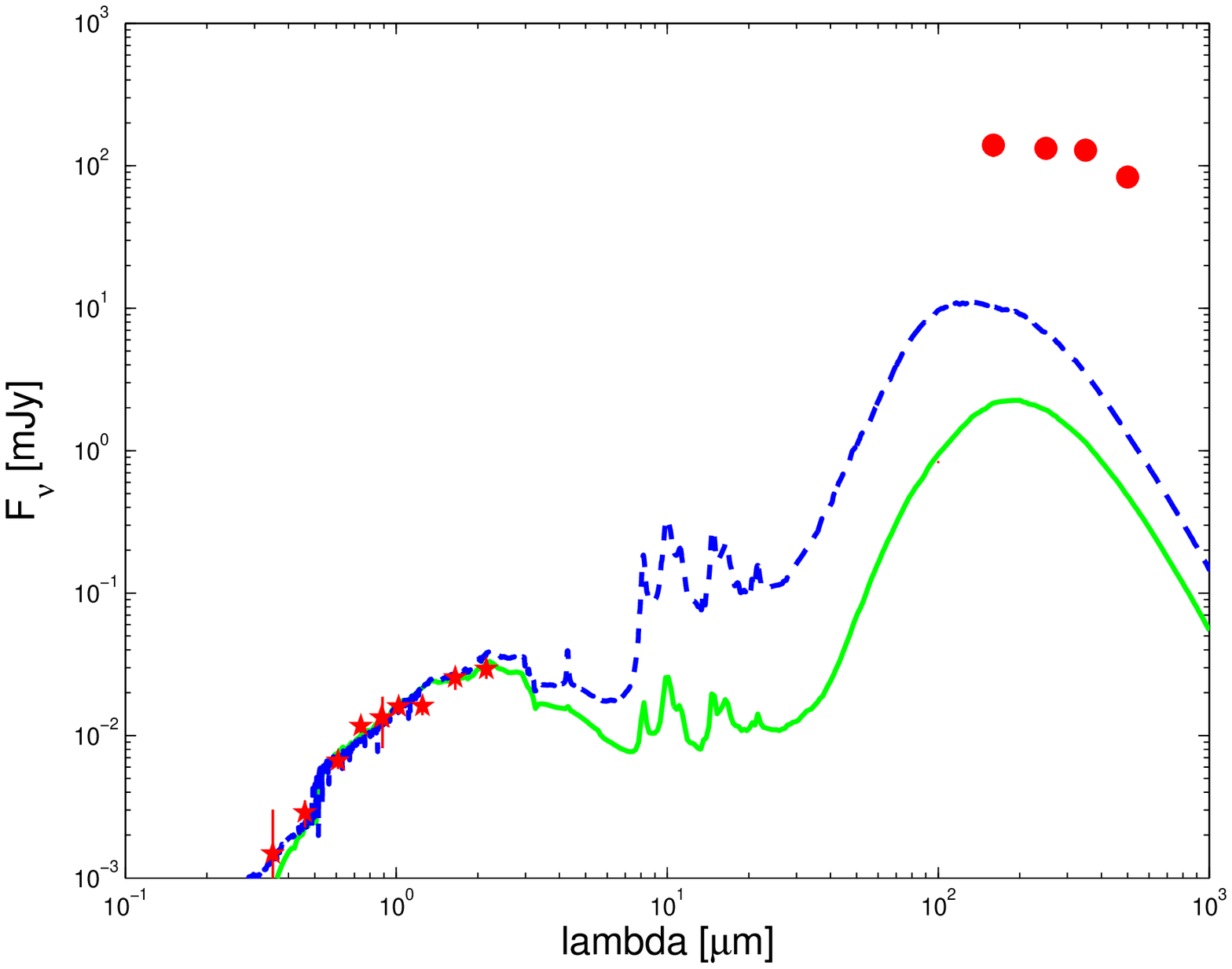}
  \includegraphics[width=0.45\textwidth]{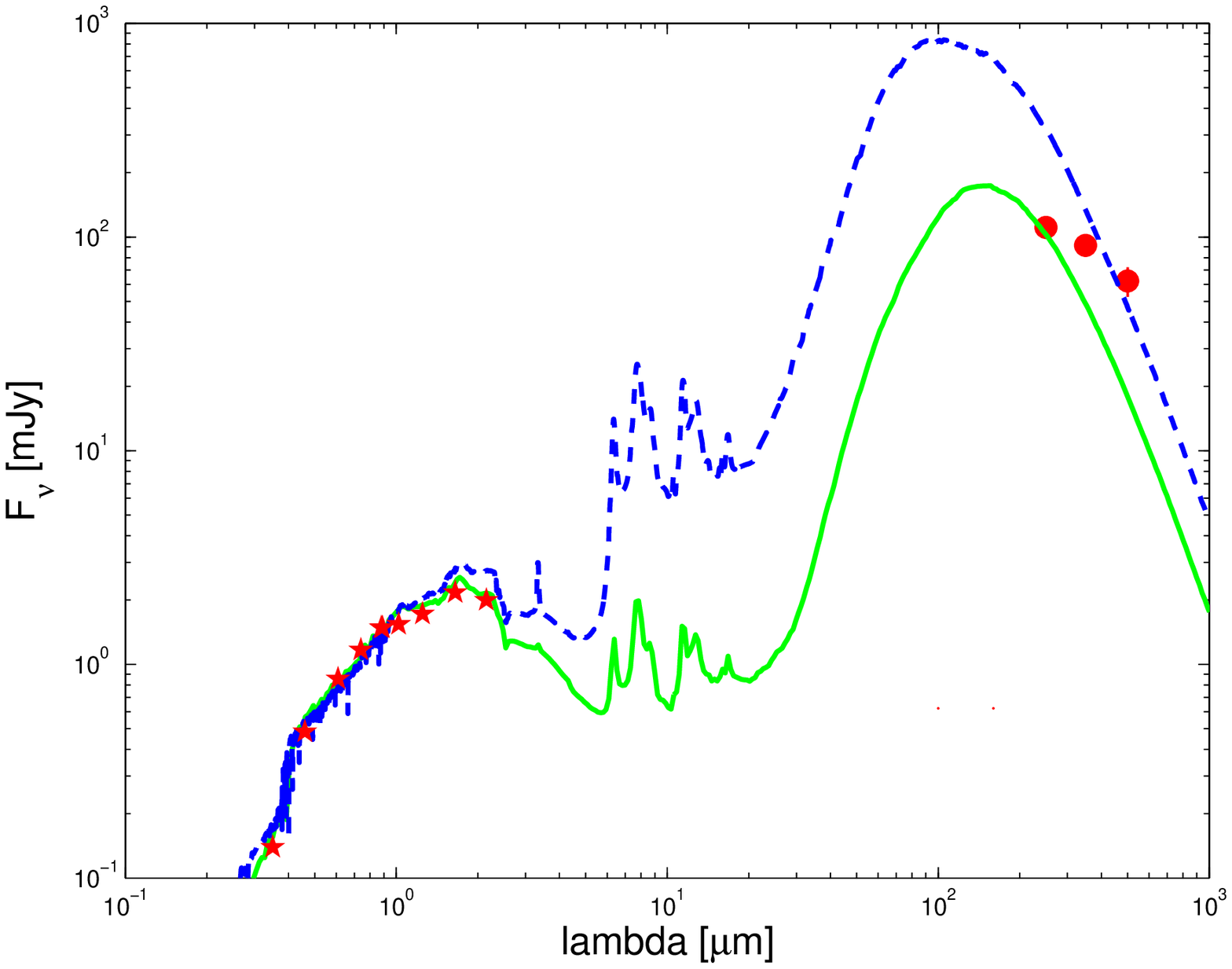}
\caption{Mean (solid green line) and high far-IR luminosity (dashed blue line) SEDs of low-$z$ H-ATLAS galaxies \citep{Smith11b} fitting the optical/near-IR (SDSS and VIKING) photometry of the optical counterparts of SDP.72 (left) and SDP.128 (right). SDP.72 is an example of objects whose far-IR to optical luminosity ratios are too large to be accounted for by a cold far-IR SED of the kind observed for $z\lsim 0.5$ galaxies with moderate SFRs, while SDP.128 is an example of objects that may be at low $z$ even though they passed the color selection described in \S\,\protect\ref{sec:sample}. The three objects of the latter kind were dropped from our sample of candidate strongly lensed galaxies. On the contrary, the optical counterparts of objects of the former kind are likely to be foreground galaxies that may act as gravitational lenses.}
\label{fig:rcheck_id72_128}%
\end{center}
\end{figure}

\begin{figure}
\begin{center}
\includegraphics[width=1\textwidth]{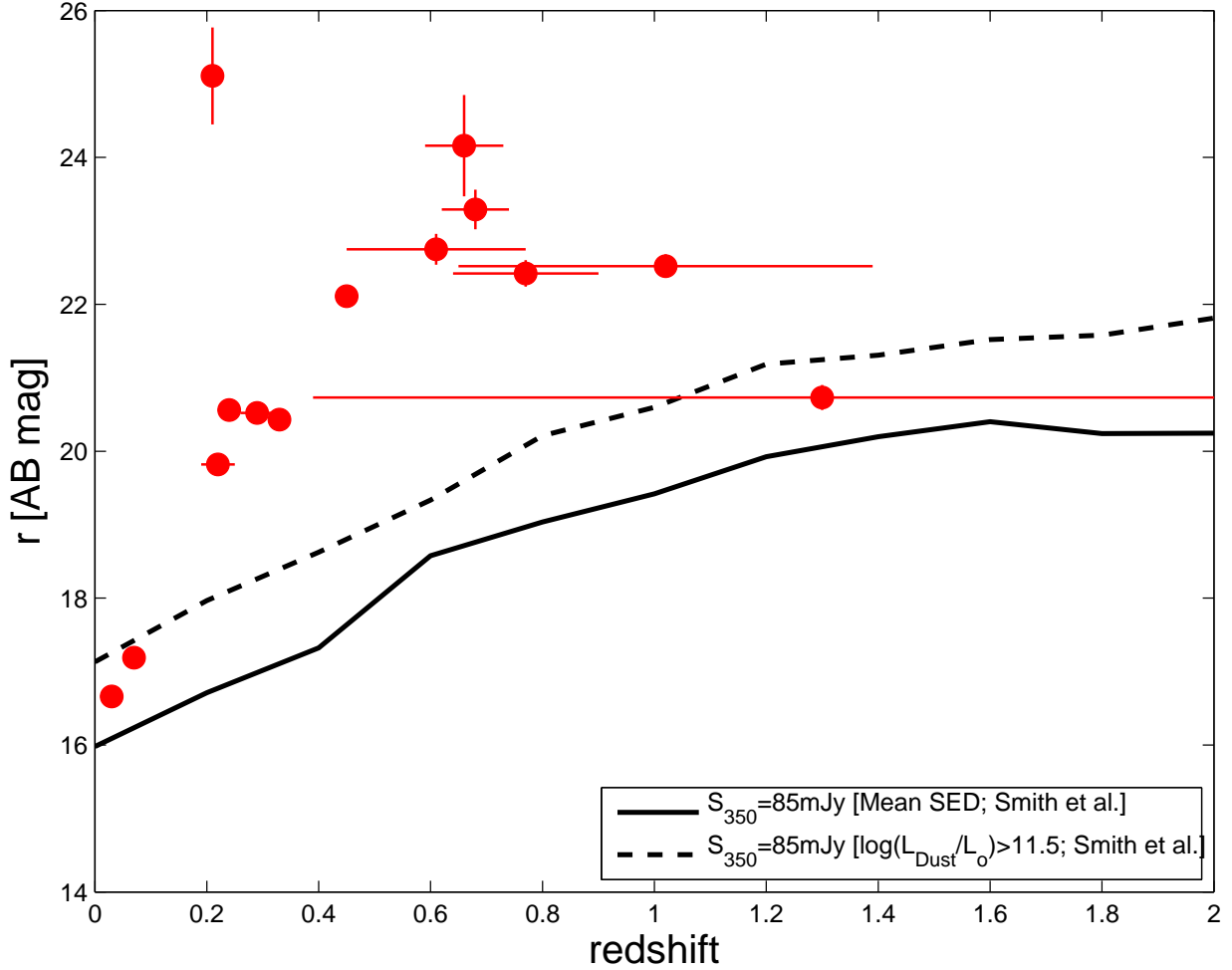}
\caption{Optical (SDSS $r$-band) magnitudes as a function of redshift for several
SEDs, normalized to $S_{350}$ = 85mJy, the flux density detection limit of our sample. The black curves refer to the mean SEDs of optically identified $z < 0.5$ SDP galaxies (solid) and to the mean SED of those in the highest
[$11.5 < {\rm log}(L_{\rm dust}/L_{\odot}) < 12$] luminosity bin (dashed) of \citet{Smith11b}. Red filled circles refer to objects with reliable ($R>0.8$) SDSS counterparts according to \citet{Smith11}; data are taken from that paper. Only for three objects the data may be compatible with the optical counterparts being the genuine identifications. They have been conservatively removed from the sample.}
\label{fig:rcheck}%
\end{center}
\end{figure}

Of the remaining 67 objects, 14 have reliable ($R>0.8$) SDSS counterparts according to \citet{Smith11}. Four of these 14 objects are strongly lensed sources of \citet{Negrello10}\footnote{The fifth Negrello source does have an SDSS counterpart but its r-band magnitude is above the limit adopted in \citet{Smith11} for estimating the reliabilities.}. Their counterparts as well as those of another seven of the 14 objects have SDSS magnitudes too faint to account for the optical and the far-IR emissions at the same time {\it if they have the cold far-IR SEDs observed for $z\lsim 0.5$ galaxies with moderate SFRs} \citep[][see the left-hand panel of Fig.~\ref{fig:rcheck_id72_128} and Fig.~\ref{fig:rcheck}; other examples are in \protect\citet{Negrello10}]{Smith11b}. 
In other words, the H-ATLAS objects must have higher apparent far-IR to optical luminosity ratios than the \citet{Smith11} galaxies, akin to those of SMGs, and/or have colder far-IR colors, and this implies that they must be at higher redshifts than those indicated by the optical/near-IR SEDs of their SDSS counterparts. We therefore assume that the SDSS counterparts are not the optical
identifications of the far-IR sources and are instead the lenses. The seven objects are HATLAS J091331.3-003642 = SDP.44, HATLAS J090952.9-010811 = SDP.60, HATLAS J090957.6-003619 = SDP.72, HATLAS J091351.7-002340 = SDP.327, HATLAS J090429.6+002935 = SDP.354, HATLAS J090453.2+022018 = SDP.392, and HATLAS J085859.2+002818 = SDP.512.

In general, SDSS counterparts that are not the optical identifications of the far-IR objects can contaminate the \textit{Herschel} photometry, but only marginally if they have the far-IR SEDs observed for $z\lsim 0.5$ galaxies with moderate SFRs (see, e.g., the left-hand panel of Fig.~\ref{fig:rcheck_id72_128}). The possible contamination affects mostly the shortest \textit{Herschel} wavelengths and may thus make the observed SEDs slightly bluer than those of lensed sources, leading to an underestimate of their photometric redshifts.

For the remaining three objects out the 14 with SDSS counterparts (HATLAS J090244.7 +013325 = SDP.112, HATLAS J091051.1+020121 = SDP.128, and HATLAS J090050.9+ 010942 = SDP.165) the data may be compatible with the optical counterparts being the genuine identifications and with them being at $z<1$ (one example is shown in the right-hand panel of Fig.~\ref{fig:rcheck_id72_128}). These objects have been conservatively removed from the sample, although further investigation may confirm some of them as valid SLG candidates.

The other 64 galaxies with $z_{\rm phot, source}\ge 1.2$,  $S_{350}\ge 85\,$mJy and $S_{250}\ge 35\,$mJy are listed in Table~\ref{tab:slg_list}. They constitute our parent sample of very bright, high redshift galaxies, among which we will search for the candidate SLGs. With this sample we also re-assess the bright end of the high-$z$ far-IR luminosity function, as discussed in the next section.

\section{The bright tail of the high-$z$ far-IR luminosity function}\label{sec:LF}

\subsection{Far-IR photometric redshifts}\label{sec:photoz}

We estimate the redshifts of objects in our parent sample in the same way as \citet{Lapi11}. The redshift estimate is the result of a minimum $\chi^2$ fit of each of the SED templates (SMM J2135-01012, Arp220, G15.141) to the SPIRE and PACS (which are mostly upper limits) data. Possible effects that could introduce a bias in our photometric redshifts are discussed in \citet{Lapi11}.

\begin{figure}
\begin{center}
\includegraphics[width=1\textwidth]{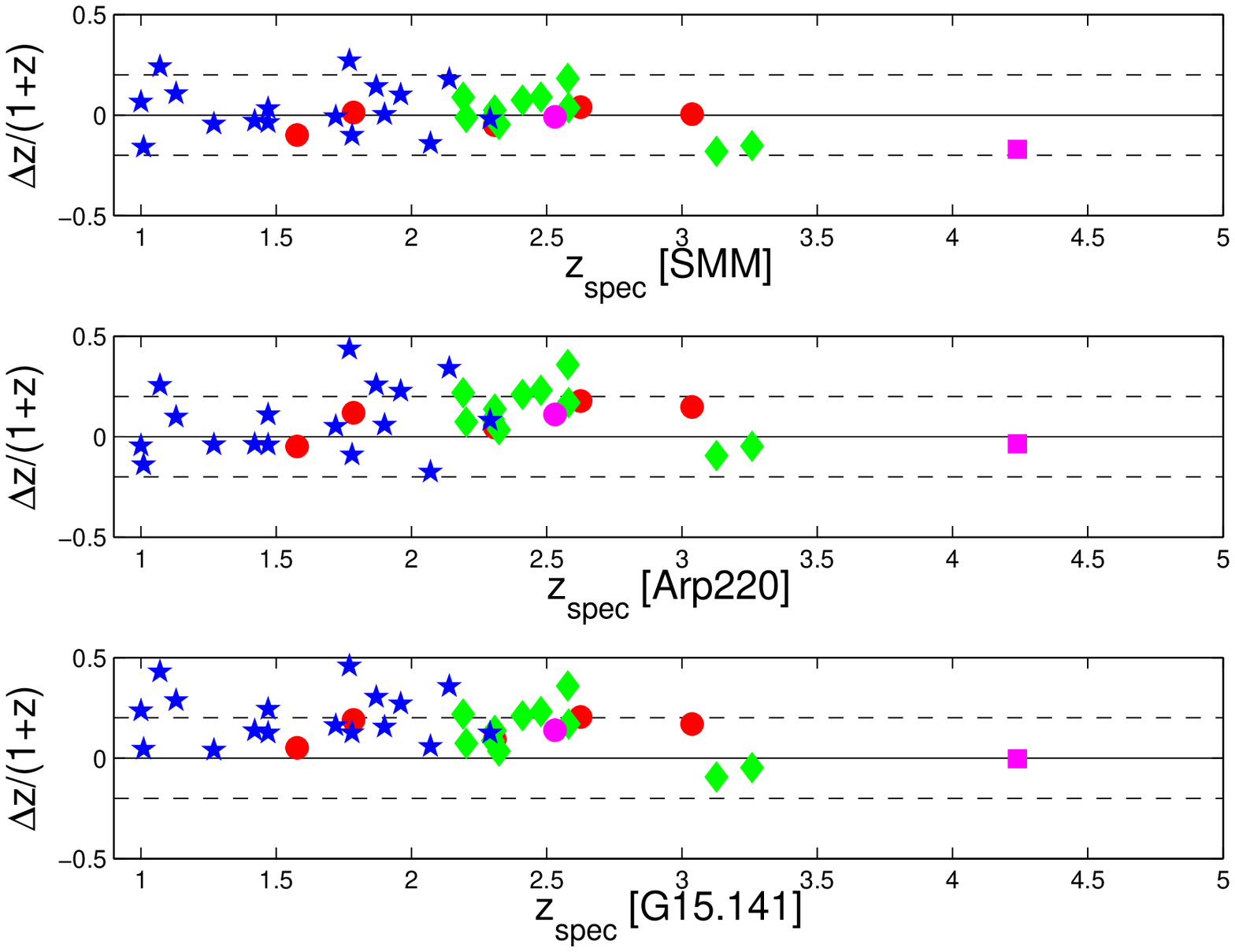}
\caption{Comparison of the photometric redshifts obtained fitting the SEDs of SMM~J2135$-0102$ (top panel), Arp220 (central panel) and G15.141 (bottom panel) to the SPIRE data for objects at $z>1$ with available spectroscopic redshifts: \citet[][ blue stars]{Bonfield11}, \citet[][ red circles]{Negrello10}, \citet[][ green diamonds]{Harris12}, \citet[][ magenta square]{Cox11}, and \citet[][ magenta circle]{Massardi11}. The dashed lines correspond to $|\Delta z/(1+z)|\equiv |(z_{\rm phot}-z_{\rm spec})|/(1+z_{\rm spec})=0.2$.}
\label{fig:photoz}%
\end{center}
\end{figure}

In Fig.~\ref{fig:photoz} we compare our photometric redshift estimates with spectroscopic measurements for the 36 H-ATLAS galaxies at $z\gtrsim 1$ for which spectroscopic redshifts are available. There is no indication that photometric redshifts are systematically under- or over-estimated when we use the SED of SMM J2135$-0102$ as a template. The median value of $\Delta z/(1+z)\equiv (z_{\rm phot}-z_{\rm spec})/(1+z_{\rm spec})$ is $-0.002$ with a dispersion of 0.115 and, remarkably, there are no outliers. These values are close to, or slightly better than those found by \citet{Lapi11} with fewer spectroscopic redshifts (24 rather than 36). The situation is only moderately worse in the case of Arp220: the median value of $\Delta z/(1+z)$ is 0.093 with a dispersion of 0.150. The median offset between photometric and spectroscopic redshifts increases to 0.158, with a dispersion of 0.124, if we use the cooler SED of G15.141. The three templates gives fits with similar $\chi^2$ values implying that it is not possible to further improve the photometric redshift precision without additional information.

As in \citet{Lapi11} we adopted the SED of SMM J2135-0102 as our reference.

\subsection{Far-IR luminosity functions}\label{subsec:LF}

We have computed the contributions of the 64 galaxies of our parent sample  to the luminosity functions at the rest-frame wavelength of $100\,\mu$m in same redshift intervals as in \citet{Lapi11}, namely $1.2\le z_{\rm source} < 1.6$, $1.6\le z_{\rm source} < 2$, $2\le z_{\rm source} < 2.4$, and $2.4\le z_{\rm source} < 4$. To do so we exploit the classical \citet{Schmidt68} $1/V_{\rm max}$ estimator, together with redshift estimates and $K$-corrections computed with the reference SED (SMM J2135-01012; see Fig.~\ref{fig:LF}). The upper scale in this figure displays the SFR corresponding to the $100\,\mu$m luminosity for the SMM J2135-0102 calibration giving
\begin{equation}
{L_{100\mu\rm m}\over {\rm W~Hz}^{-1}}=5.9\times 10^{23}\,{\mathrm{SFR}\over
M_{\odot}~{\rm yr}^{-1}},
\end{equation}
\citep[][assuming a \citet{Chabrier03} initial mass function]{Lapi11}. Since for galaxies with intense star formation the rest-frame dust emission peaks in the range $\lambda\approx 90-100\, \mu$m, the $100\, \mu$m luminosity is good estimator of the SFR.

Our estimates join smoothly with those of \citet{Lapi11} at the lowest apparent (i.e. uncorrected for the effect of gravitational lensing) luminosities but show an indication of a flattening at the highest apparent luminosities. This flattening is expected as the effect of strong lensing, in analogy to what happens with the source counts. It was not present in \citet{Lapi11} because all objects with SDSS counterparts were removed from their sample. Strongly lensed galaxies unavoidably dominate the highest apparent luminosity (or flux density) tail of the observed luminosity functions (or number counts), where the space (or surface) density of unlensed galaxies drops very rapidly. The flattening induced by these objects reflects the flatter slope of the sub-L$_\ast$ luminosity function (or the flattening of faint counts). The five strongly lensed galaxies identified by \citet{Negrello10} fall on this part of the luminosity function.

Although there is a clear analogy between the behaviour of the luminosity functions and that of the source counts, the latter are integrated quantities. As a consequence, even in the case of modestly accurate photometric redshifts, the luminosity functions in redshift bins are a much stronger discriminator of strongly lensed galaxies than the number counts in flux density bins. This is the basis of our approach for extending the selection of candidate SLGs to fainter flux densities.

\section{Identification of SLG candidates in the SDP area}\label{sec:optcrit}

\subsection{Optical/near-IR counterparts}

An important ingredient for our selection of candidate SLGs is the close association with a galaxy that may qualify as the lens. As we have seen, only 11 galaxies in our parent sample have such an association in the SDSS. The VIKING survey drastically improves the situation and indeed turns out to be well-suited for our purpose. VIKING is one of the public, large-scale surveys ongoing with VISTA, a 4-m class wide-field ESO telescope situated at the Paranal site in Chile \citep{Emerson10}. It aims at covering around 1500 deg$^2$ of the extragalactic sky, including the GAMA 9h, 12h, and 15h fields, plus both H-ATLAS South Galactic Pole (SGP) fields, in 5 broad-band filters, $Z$, $Y$, $J$, $H$, and $K_s$. The median image quality is $\approx 0.9$ arcsec, and typical $5\sigma$ magnitude limits are $J \approx 21.0$, and $K_{\rm s} \approx 19.2$ in the Vega system.

We matched our objects with the preliminary object catalogues of the VIKING survey in the GAMA 9h field \citep{Fleuren11} within a search radius of $10''$. We found 106 possible VIKING counterparts to 58 of our 64 objects ($\sim 91\%$). When there is more than one possible counterpart, we selected the one closest to the SPIRE position, that frequently coincides with the highest reliability ($R$) counterpart, as determined by Fleuren et al. (2011); $\sim 53\%$ of them have $R>0.8$. Note that $R<0.8$ does not mean that the object is not an identification or a lens.

Figure~\ref{fig:angdist} shows how the ratio of the mean surface density of matched VIKING galaxies to their overall mean density varies with the angular distance from objects in our parent sample. There is  a clear overdensity for radii smaller than 3.5 arcsec, indicating a high likelihood of some physical relation between the VIKING and the sub-mm object: they may either be the same object or be related by lensing. Since 3.5 arcsec is roughly the angular distance between the lens and the lensed images where the separation distribution drops \citep{Kochanek06} we have selected as our primary candidate SLGs the H-ATLAS objects with a VIKING association within 3.5 arcsec, i.e. 34 objects. The object HATLAS J090739.1-003948 = SDP.639 has two close optical counterparts, one of which has a photometric redshift ($z_{\rm phot} \simeq 2.62\pm0.4$), estimated by us, compatible with our photometric redshift of the H-ATLAS source ($z_{\rm phot, source}=2.89\pm0.4$), and may thus be the identification of the lensed source, while the second has $z_{\rm phot}=0.39\pm0.15$, and may be the lens. 

The VIKING survey has provided the $Z-H$ colors for 19 candidate lenses in our sample (see Fig.~\ref{fig:optcolor}). With 5 exceptions (HATLAS J090957.6-003619 = SDP.72, HATLAS J090626.6+022612 = SDP.132, HATLAS J090931.8+000133 = SDP.257, HATLAS J090950.8+000427 = SDP.419, and HATLAS J090739.1-003948 = SDP.639) the colors of candidate lenses are consistent with them being passive early-type galaxies.

The counterpart of the object SDP.180 (HATLAS J090408.6+012610) has photometric data in only two bands ($J$ and $K_s$). Since we could not determine whether or not it may be the true identification, we have conservatively dropped it from our sample of candidate lenses. This leaves us with 33 objects that have candidate VIKING lenses.

\subsection{Photometric redshifts of candidate VIKING lenses}

A substantial fraction of VIKING associations to H-ATLAS objects have either spectroscopic or (in most cases) photometric redshifts \citep{Fleuren11}. The latter were obtained with the publicly available code ANNz \citep{Collister04}, combining the VIKING near-infrared photometry with the optical photometry from the SDSS. Above $z\sim 0.8$, where the training set for the neural networks used by ANNz is less rich, the code frequently fails to converge to a solution. This happens for a negligible fraction of galaxies in the whole VIKING catalogue but for a large fraction of our 33 objects (see Table \ref{tab:slg_list}). For the 23 objects in the parent sample without ANNz redshift and for 3 additional objects for which ANNz formally converges but gives exceedingly large errors (see Table \ref{tab:slg_list}) we have made our own photometric redshift estimates. To this end we have used a library of 16 SEDs of early-type galaxies\footnote{Four out of the 5  galaxies with blue $Z-H$ colors have ANNz redshifts. The fifth has colours not far from those of an early-type galaxy, and redder than those of late-type galaxies.}, computed with GRASIL\footnote{http://galsynth.oapd.inaf.it} \citep{Silva98} with updated stellar populations. We ran two chemical models suitable for a typical early type galaxy, i.e. with an efficient star formation rate for the first Gyr, and passively evolving thereafter. The two chemical evolution models have a different
metal enrichment history with a SFR-averaged metallicity $\langle Z\rangle
\sim 0.05$ and $\langle Z\rangle \sim 0.03$, respectively. For each chemical
evolution model we have then computed a series of synthetic SEDs at eight
selected ages, between 2 Gyr and 9 Gyr.

For each SED template the redshift was estimated through a minimum $\chi^2$ fit of the SDSS (available only for 5 of the 23+3 objects) and VIKING photometric data (including upper limits). The adopted photometric redshift is the median value obtained with the different SEDs and the associated error is the rms difference from the median (typically $\sim0.1$)\footnote{For those objects with both ANNz and our ($z_{\rm phot}$) redshift estimates the median value of $|\Delta z/(1+z)|\equiv |(z_{\rm phot}-z_{\rm ANNz})| /(1+z_{\rm ANNz})$ is $0.09$ with a dispersion of 0.16.}. Note that, as illustrated by Fig.~\ref{fig:optcolor}, the adoption of a late-type SED template would generally have implied much higher photometric redshifts (because a late-type galaxy can become that red only at high $z$) and, hence, extreme stellar masses.

Both theoretical expectations for objects in the redshift range considered here (see, e.g., Fig.~\ref{fig:zopt}) and observational data from surveys with a \textit{\rm source} redshift distribution similar to ours, namely the Cosmic Lens All-Sky Survey  \citep[CLASS, ][]{Browne03}, the SDSS Quasar Lens Search \citep[SQLS,][]{Oguri06, Oguri08}, and the COSMOS survey (\citealt{Faure08, Jackson08}; see Fig.~7 of \citealt{Treu10}), indicate a cut-off at $z_{\rm lens}\simeq 0.2$ in the redshift distribution of \textit{lenses}. Although the observed cut-off may be, at least partly, due to an observational bias (at low $z_{\rm lens}$ the lens galaxies are brighter and more extended, and may therefore confuse the images) we have conservatively dropped from our sample of strong candidate SLGs the two objects whose candidate deflectors are at $z<0.2$ (in any case, they are highlighted in Table~\ref{tab:slg_list} for follow-up purposes). An estimate of the ``lens probability'' for these objects is given in \S\,\ref{sec:purity}.

\subsection{Sample purity}\label{sec:purity}

Although we have been as conservative as possible in the selection of the candidate SLGs, some contamination of the sample is unavoidable. First, given the wide variety of galaxy SEDs, we cannot be absolutely sure that the VIKING counterparts of {\it all} our 31 strong candidates are foreground galaxies and not the identifications of sources themselves. However, only in a minority of cases (ID\,9, 11, 53, 79, 122, 309) we find that with a fine tuning of parameters controlling the star-formation history and the dust re-emission spectrum we can roughly account for both the optical/near-IR and the {\it Herschel} photometric data assuming that they refer to the same source. But two of these galaxies (ID\,9 and 11) were already shown by Negrello et al. (2011) to be strongly lensed. In these cases the lens galaxy was clearly identified and was found to have near-IR magnitudes close to those of the source. This suggests that some ambiguous cases can be misinterpreted in either direction: in a few cases VIKING counterparts interpreted as foreground lenses may be the genuine identifications of the sources; in other cases alleged identifications may be foreground lenses. Since also for the other 4 objects the interpretation of all the photometric data as referring to a single source is intricate and the fit is anyway poor, we have decided to keep them in our sample.

Second, given the uncertainties on the redshifts of candidate sources and lenses, on the mass and density profiles of the candidate lenses and, especially, on source positions, even if the VIKING sources are foreground galaxies they may not yield a strong (i.e. a factor of at least 2) gravitational amplification. The typical positional uncertainty of H-ATLAS sources with $5\sigma$ detections at $250\,\mu$m is $\simeq 2.4''$, and decreases proportionally to $1/(\hbox{S/N})$ \citep{Rigby11}. With only one exception, all our strong candidates are detected at $250\,\mu$m with $\hbox{S/N}\ge 10$ and their positional errors are therefore $\le 1.2''$. For each source we have computed tentative estimates of the ``lens probability'', i.e. of the probability of a ``strong'' gravitational amplification ($\mu \ge 2$). To this end we have adopted the photometric or measured redshifts of the candidate source and lens, have estimated the halo mass using $M_\star/L_K=1$ \citep{Williams2009} and $M_{\rm halo}/M_\star$ from \citet{Moster2010} and \citet{Shankar2006}, and have used a Single Isothermal Sphere (SIS) profile for the lens. The distribution of angular separations between the source and the VIKING counterparts was modeled as a Gaussian with mean equal to the nominal separation and dispersion $\sigma_{\rm sep}=2.4''[5/(\hbox{S/N})]$.  The lens probability was then obtained as the area of the Gaussian over the range of angular separations yielding $\mu \ge 2$.  We define the ``purity'' of the sample, as a function of $S_{350}$ or of the angular separation, as the ratio between the sum of lens probabilities and the number of lensed candidates within each flux density or angular separation bin. The results are displayed in Fig.~\ref{fig:purity}. As expected the ``purity'' declines with increasing angular separation and with decreasing flux density. The global ``purity'' of the sample is $72\%$.

Although the ``lens probabilities'' of individual objects are quite uncertain and should therefore be used mainly for statistical purposes, e.g. to estimate the sample purity as made above, our analysis has picked out 4 objects, identified in Table~\ref{tab:slg_list}, whose lensing probability is particularly low ($<30\%$ and down to $\simeq 1\%$) and therefore are unlikely to be strongly lensed. Two of these objects (SDP.72 and SDP.257) have blue colors, indicative of a late-type galaxy; the other two (SDP.98 and SDP.290) are close to the adopted limit on the angular separation between the candidate source and the candidate lens (angular separation $\ge 3''$). For the two objects whose candidate deflectors are at $z<0.2$, and were excluded from the sample, the lensing probabilities are 0.4\% (SDP.545) and 24\% (SDP.354), confirming that they are not good SLG candidates.

\subsection{Redshift distributions}

The global redshift distributions of the 31 SLG candidates and of the associated lenses are shown in Fig.~\ref{fig:zopt}. In the same figure we also show that dropping the 8 objects with the lowest estimated lens probabilities (probabilities $<50\%$) does not substantially change the shape of the redshift distributions. Figure~\ref{fig:treu} compares the redshift distribution of our lens candidates with those of the  CLASS, COSMOS, SLACS \citep{Auger09}, and SQLS surveys, as given in Fig.~7 of \citet{Treu10}. The main difference is that our lens candidates are found out to much higher redshifts than those of the other surveys. If confirmed (and the agreement with theoretical expectations is quite reassuring in this respect), this result implies that our selection allows one to substantially extend the redshift range over which gravitational lensing can be exploited to study the lens galaxy structure and its evolution. We note, in particular, that there is observational evidence of a strong size evolution of massive early type galaxies from $z\sim 1$ (e.g. \citealt{Trujillo11} and references therein). The interpretation of this evolution is still controversial  however (e.g., \citealt{Oser11}; \citealt{vanDokkum10}; \citealt{Fan08, Fan10}). Different models imply different predictions for the evolution of gravitational potential in the inner parts of the galaxies; gravitational lensing will provide a test for such predictions.

Most of the candidate lenses show an excess, mainly in the $K_{\rm s}$ band, that can be attributed to the contribution of the background source, as directly seen in the HST images of bright strongly lensed galaxies \citep{Negrello12}. This effect needs to be taken into account in the photometric redshift estimates of the candidate lenses, e.g. decreasing the weight of the $K_{\rm s}$ magnitude. Whenever accurate multi-band photometry is available the $K_{\rm s}$ point was ignored altogether in our photometric redshift estimates.

Our 31 SLG candidates are emphasized in boldface in Table~\ref{tab:slg_list}, where we have also identified the 4 objects for which we have tentatively estimated a lens probability $<30\%$. The SLG candidates have high apparent luminosity (Fig.~\ref{fig:LF}), high redshift (Fig.~\ref{fig:zopt}, top panel), and are associated with foreground galaxies, mostly with near-IR colors of early-type galaxies (Fig.~\ref{fig:optcolor}), at angular separations $\le 3.5$ arcsec (Fig.~\ref{fig:angdist}). The five confirmed SLGs from \citet{Negrello10} are among our best candidates.

\section{Selecting candidate SLGs from SPIRE data alone}\label{sec:lumsel}

As discussed above, the VIKING data play a key role in the selection of our SLG candidates, since they allow us to identify the associated candidate lenses. Although the VIKING survey plans to cover $\simeq 1500\,\hbox{deg}^2$, it will not cover the H-ATLAS North Galactic Pole field, and near-IR surveys to the same depth of the missing areas are not foreseen. However, a high efficiency selection of candidate strongly lensed sources fainter than $S_{500}=100\,$mJy is possible using only SPIRE data. This is readily apparent from the previous results: almost 50\% of objects selected with the criteria $S_{350}\ge 85\,$mJy, $S_{250}\ge 35\,$mJy, $S_{350}/S_{250}>0.6$, and $S_{500}/S_{350}>0.4$ turn out to be strong candidate SLGs, in close agreement with the predictions of the \citet{Lapi11} model. This is already an impressively high fraction, especially in consideration of how easily it is achieved. However, the selection efficiency can be further improved by exploiting the fact that SLGs dominate the highest apparent luminosity tail of the high-$z$ far-IR/sub-mm luminosity function (Fig.~\ref{fig:LF}).

To investigate the potential of an approach relying only on {\it Herschel}/SPIRE photometry, we have computed, using the SED of SMM~J2135$-0102$, the photometric redshifts of SDP objects with $S_{250}>35\,$mJy and $S_{350}$ above the $4\sigma$ limit, a sample almost completely overlapping the one defined by \citet{Lapi11}, except that the $R>0.8$ SDSS associations are not removed. The redshifts were split in bins of $\delta z_{\rm source} =0.1$, and within each bin we have selected the objects with $S_{350}>85\,$mJy, $z_{\rm source}>1.2$ and apparent luminosity above a given percentile.
In order to moderate the dependence of the results on a particular SED, we have repeated the procedure using the three SEDs discussed in \S\,\ref{sec:photoz} and consider only the candidates selected by all the SEDs. Finally we require that objects  have $S_{350}>85$ mJy and $z_{\rm source}>1.2$, like in \S\,\ref{sec:sample}. Figure~\ref{fig:lumsel} (bottom panel) shows, as a fraction of the top apparent luminosity percentile, the percentage (left-hand scale) and the number (right-hand scale) of strongly lensed candidates, as identified in \S\,\ref{sec:optcrit}. For example, about $70$\% of objects having apparent $100\,\mu$m luminosity in the top 2\% (21 objects) were previously identified as strong SLG candidates.

\section{Summary and conclusions}\label{sec:discussion}

We have presented a simple method, that will be referred to as the {\it Herschel}-ATLAS Lensed Objects Selection (HALOS), that gives the prospect of identifying roughly 1.5--2 strong SLG candidates per square degree from the H-ATLAS survey, i.e. about 1000 over the full survey area. This amounts to a factor $\simeq 4$--6 increase compared to the surface density of SLGs brighter than $S_{500}=100\,$mJy, whose selection has proven to be easy. Samples of thousands of strongly lensed systems are needed to make substantial progress on several major astrophysical and cosmological issues, as stressed by Treu (2010). Also, the extension to fainter flux densities is crucial to pick up galaxies representative of the bulk of the star-forming galaxy population at $z_{\rm source} \simeq $1--3, that without the upthrust of strong lensing are fainter than the SPIRE confusion limit.

The method appeals to the fact that strongly lensed galaxies inevitably dominate the highest apparent luminosity tail of the high-$z$ luminosity function. The first step is therefore to pick up high apparent luminosity and high-$z$ galaxies. The primary selection, based on SPIRE photometry ($S_{350}\ge 85\,$mJy, $S_{250}\ge 35\,$mJy, $S_{350}/S_{250}>0.6$, and $S_{500}/S_{350}>0.4$), has yielded a sample of 74 objects in the H-ATLAS SDP field of $\simeq 14.4\,\hbox{deg}^2$. After having rejected intruders of various types (see \S\,\ref{sec:sample}) we are left with a sample of 64 objects, with estimated redshifts $\ge 1.2$. This sample has allowed us to re-assess the brightest portion of the apparent $100\,\mu$m luminosity function in the same 4 redshift bins ($1.2 \le z_{\rm source} <1.6$, $1.6 \le z_{\rm source} <2.0$, $2.0 \le z_{\rm source} <2.4$, $2.4 \le z_{\rm source} <4.0$) of \citet{Lapi11}, whose sample was biased against SLGs because of the rejection of all objects with SDSS $R>0.8$ counterparts, some of which may be the foreground lenses. The new estimate of the luminosity function shows indications of a flattening at the highest apparent luminosities as expected, on the basis of the \citet{Lapi11} model coupled with the formalism by \citet{Perrotta02}, as the effect of the contribution of SLGs. This flattening reflects the flatter slope of the sub-L$_\ast$ luminosity function, and confirms that our approach has the potential of allowing us to investigate more typical high-$z$ star-forming galaxies.

To identify the candidate SLGs we have looked for close associations (within 3.5 arcsec) with VIKING galaxies that may qualify as being the lenses.  We found 34 such associations. 
The optical/near-IR data for 32 of these objects were found to be incompatible (or, in 4 cases, hardly compatible) with them being the identifications of the H-ATLAS objects. Another object has two close VIKING counterparts, one of which may be the identification and the other may be the lens. We kept this object as a candidate SLG. The VIKING data on the counterpart of the last of the 34 objects are insufficient to decide whether it is a likely lens, and we conservatively dropped it.

Again to be conservative we have further restricted the sample of candidate SLGs to objects whose VIKING counterparts have redshifts $\ge 0.2$ since this seems to be a lower limit to lens redshifts found in previous surveys, although there is nothing, in principle, that prevents a object at $z<0.2$ from being a lens. Thus, the 2 objects with VIKING counterparts at $z < 0.2$ should also be taken into account for follow-up observations. In this way, we end up with at least 31 high apparent luminosity and high-$z$ SDP objects, corresponding to a surface density of $\sim 2\,\hbox{deg}^{-2}$, that appear to be physically associated with foreground galaxies that are most likely the lenses.

Using the available data we have carried out, for each object, a tentative estimate of the probability that it is strongly lensed, i.e. has a gravitational amplification $\mu >2$, and of the purity of the sample as a function of the angular separation between the candidate source and the candidate lens and of the source flux density. The global purity of the sample is estimated to be 72\%. Although, given the many uncertainties, not much weight should be attached to the lens probabilities of individual objects, we have picked out 4 cases with particularly low lens probabilities ($<30\%$). 

The number counts of candidate SLGs, corrected for the flux-density dependent purity, are shown in  Fig.~\ref{fig:SNC}, where model predictions are also plotted for comparison; the agreement is good. The model indicates that the counts of candidate SLGs with $S_{350}>85$ mJy are mostly contributed by amplifications $\mu\ge3$.

As pointed out by \citet{Treu11} $\sim90$\% of the lenses discovered by SLACS are massive early-type galaxies \citep{Auger09}: only 10 of the 85 SLACS lenses have visible spiral morphology. While our approach with HALOS is completely different from the one used in SLACS and SWELLS, it is interesting that we obtain a similar ratio between blue (likely late-type) and red (likely early-type) lens galaxies. 

Excluding the candidate SLGs from the initial sample, we can constrain the bright end of the luminosity function of unlensed galaxies, which turns out to be extremely steep, as expected if these galaxies are indeed proto-spheroidal galaxies in the process of forming most of their stars in a single gigantic starburst \citep{Granato04, Lapi11}. 

The estimated redshift distribution of our {\it candidate lensed galaxies} extends up to $z_{\rm source}\simeq 3.2$ and is similar to those of other searches for strongly lensed sources, like CLASS \citep{Browne03}, SQLS \citep{Oguri06, Oguri08}, and COSMOS \citep{Faure08, Jackson08}. On the other hand, our {\it lenses} are found up to $z_{\rm lens}\simeq 1.6$--1.8 (with a peak at $z_{\rm lens}\simeq 0.8$), while in the case of the other surveys they are confined to $z_{\rm lens}<1$. We caution however that the redshift estimates are photometric, and need to be confirmed by spectroscopic measurements. If this lens redshift distribution will be validated, our selection will allow us to substantially extend the redshift range over which gravitational lensing can be exploited to study the lens galaxy mass and structure, and their evolution. In this respect, it is reassuring that the observed redshift distribution appears consistent with expectations from the model (see Fig.~\ref{fig:zopt}).

The five brightest sources among the 31 best SLG candidates were already shown to be strongly lensed galaxies through an intense multi-instrument observational campaign \citep{Negrello10}. As for the fainter ones, we envisage a  follow-up strategy comprising several steps. First we need a spectroscopic confirmation that they are at the high redshifts indicated by our photometric estimates. Millimeter-wave spectroscopy of CO transitions proved to be very effective not only for redshift measurements but also for providing dynamical information and gas masses \citep[e.g.][and references therein]{Harris12}. A comparison with expectations from the empirical relationship between CO luminosity and line-width for unlensed galaxies \citep{Bothwell12,Harris12} provides a first indication for or against the presence of gravitational amplification and, in the positive case, an estimate of its amplitude. Deep high resolution imaging is obviously necessary to establish the lensing nature of the sources by revealing and mapping the lensed images (arcs). This has been done, although for brighter sources, in the optical/near-IR \citep{Fu2012,Negrello12} and at (sub)-millimeter wavelengths \citep{Riechers2011,Bussmann12}. The latter have the great advantage that the images are little affected by, or totally immune to the effect of the lensing galaxies (which, as mentioned above, are mostly passive, early type galaxies). ALMA overcomes the problem of the limiting sensitivity of earlier (sub)-mm instruments allowing one to make very deep, high resolution images, thus making possible a detailed study of the internal structure and dynamics of the lensed galaxies.

\begin{figure}
\begin{center}
\includegraphics[width=1\textwidth]{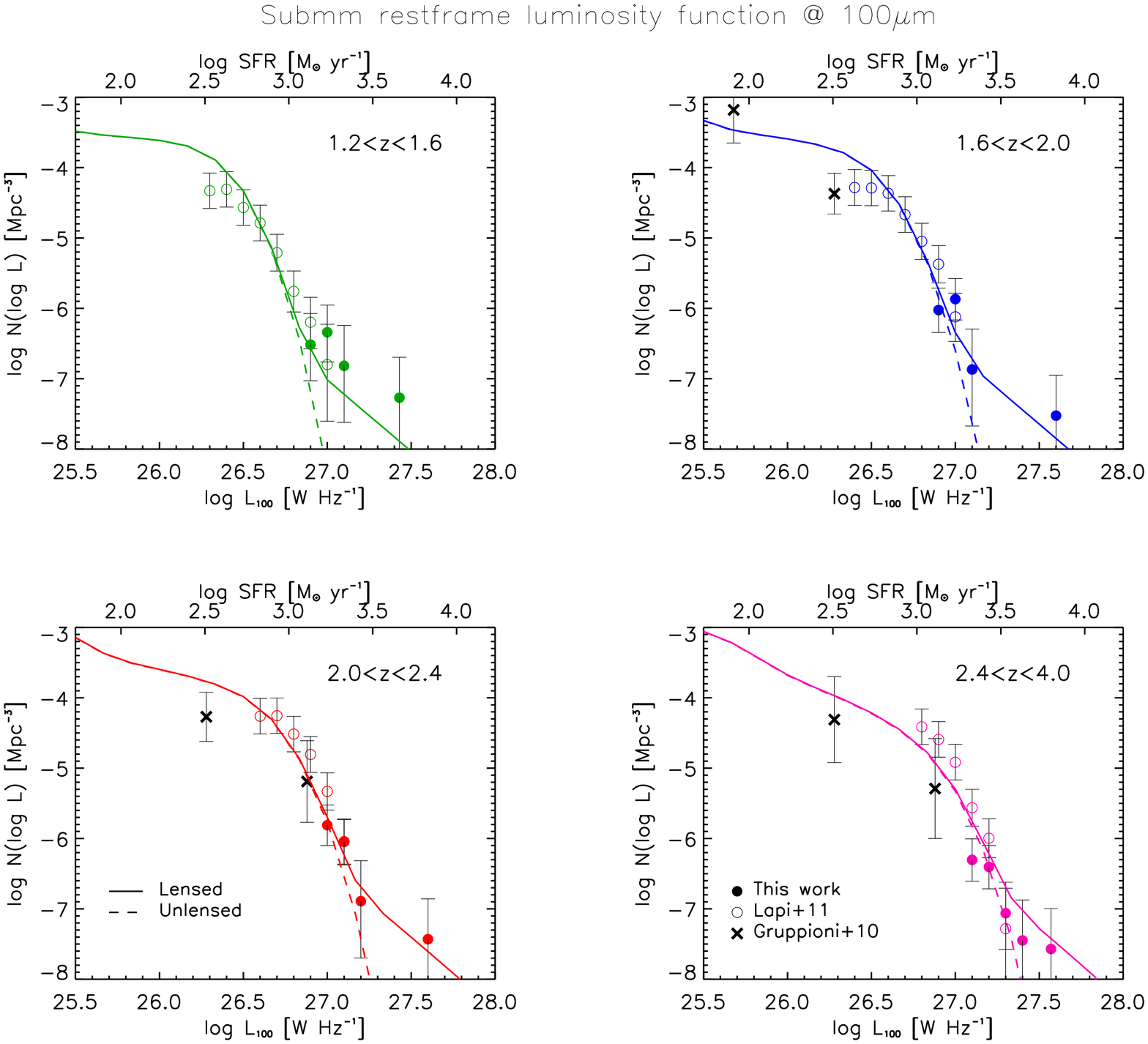}
\caption{Contributions of our bright objects (filled circles) to the $100\,\mu$m luminosity functions in different redshift intervals. The open circles show, for comparison, the estimates by \citet{Lapi11}. The crosses are $90\,\mu$m luminosity functions derived by \citet{Gruppioni10} from PACS data. The dashed lines, that coincide with the solid lines except at the highest apparent luminosities, show the model for unlensed proto-spheroidal galaxies described in \citet{Lapi11}. The solid lines include the contributions of strongly lensed galaxies (see text). }
\label{fig:LF}%
\end{center}
\end{figure}

\begin{figure}
\begin{center}
\includegraphics[width=1\textwidth]{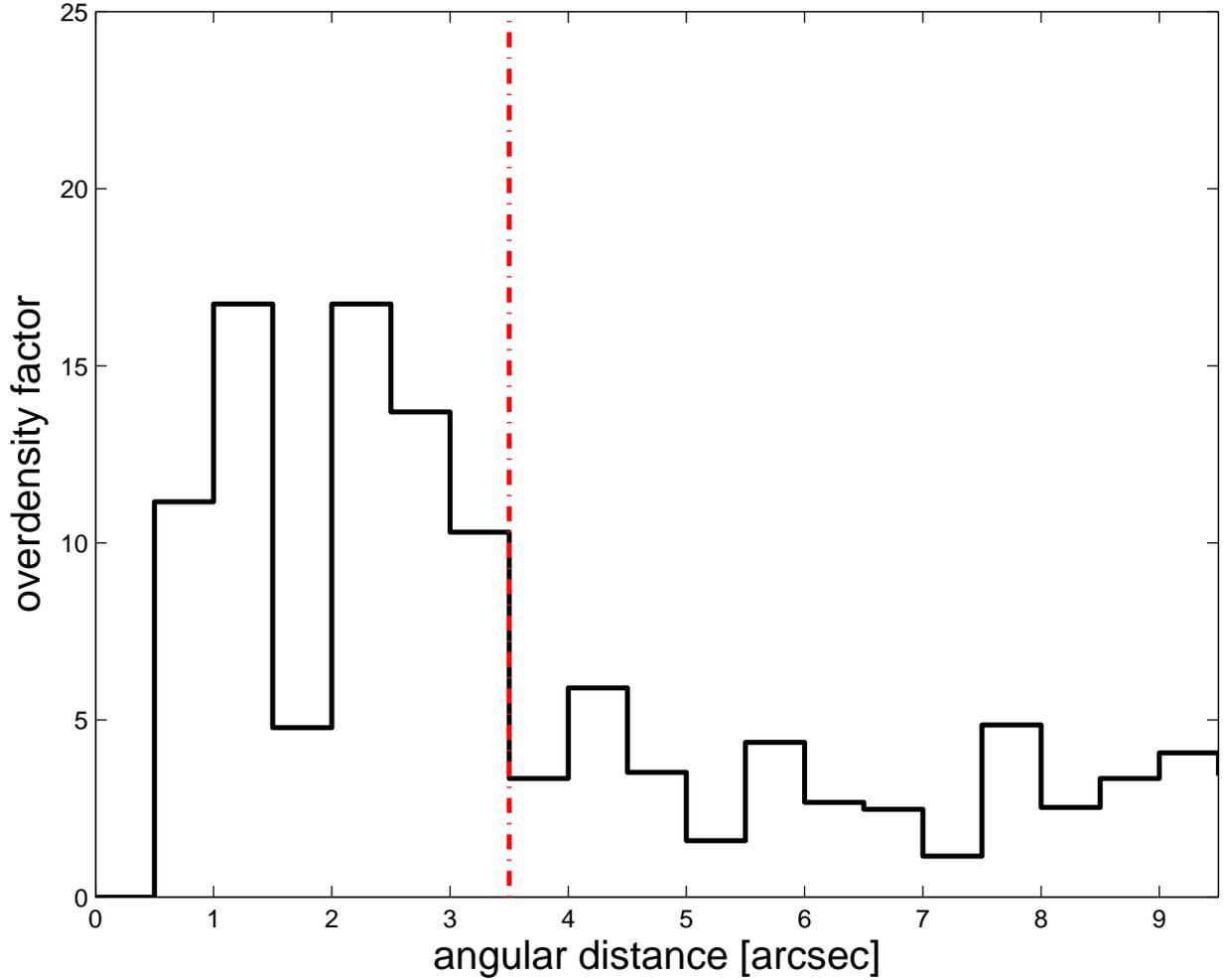}
\caption{Ratios between the mean surface densities of VIKING galaxies in annuli 0.5 arcsec wide, centered on the positions of objects in our parent sample, and the mean surface density of VIKING galaxies ($\simeq 1.19\times 10^{-3}\,\hbox{arcsec}^{-2}$) as a function of the angular distance from the objects. The vertical red dot-dashed line shows the maximum angular separation ($3.5''$) used to select our candidate SLGs.}
\label{fig:angdist}%
\end{center}
\end{figure}

\begin{figure}
\begin{center}
\includegraphics[width=1\textwidth]{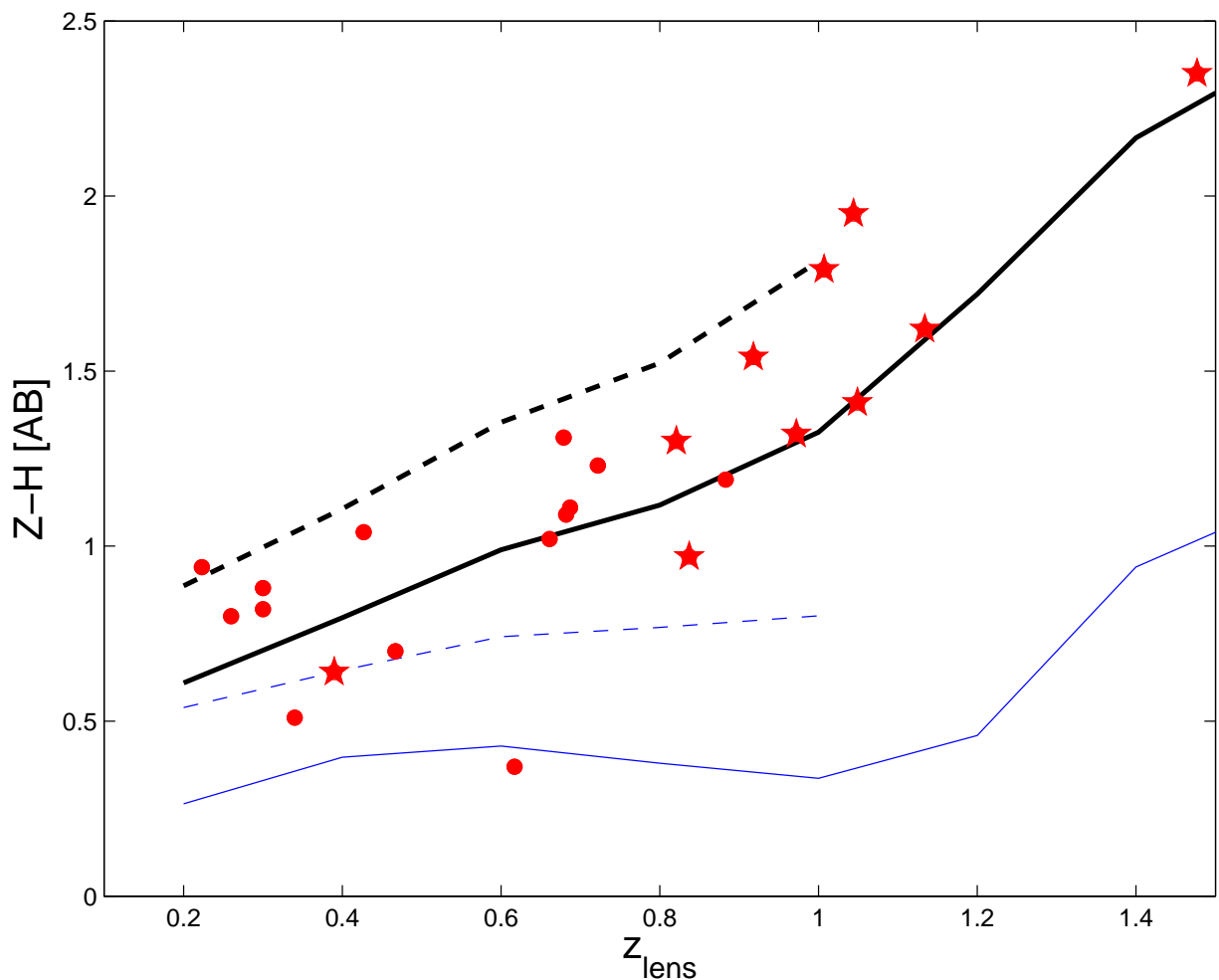}
\caption{Near-infrared ($Z-H$) color vs photometric redshift for the lens candidates (ANNz redshifts: red dots; our own estimates: red stars). The black lines are tracks of typical elliptical galaxies with ages of 3\,Gyr (solid line) or 9\,Gyr (dashed). The thin blue lines are the tracks of typical spiral galaxies with the same ages. There are 5 galaxies with colors bluer than expected for early-type galaxies.
}
\label{fig:optcolor}%
\end{center}
\end{figure}

\begin{figure}
\begin{center}
\includegraphics[width=1\textwidth]{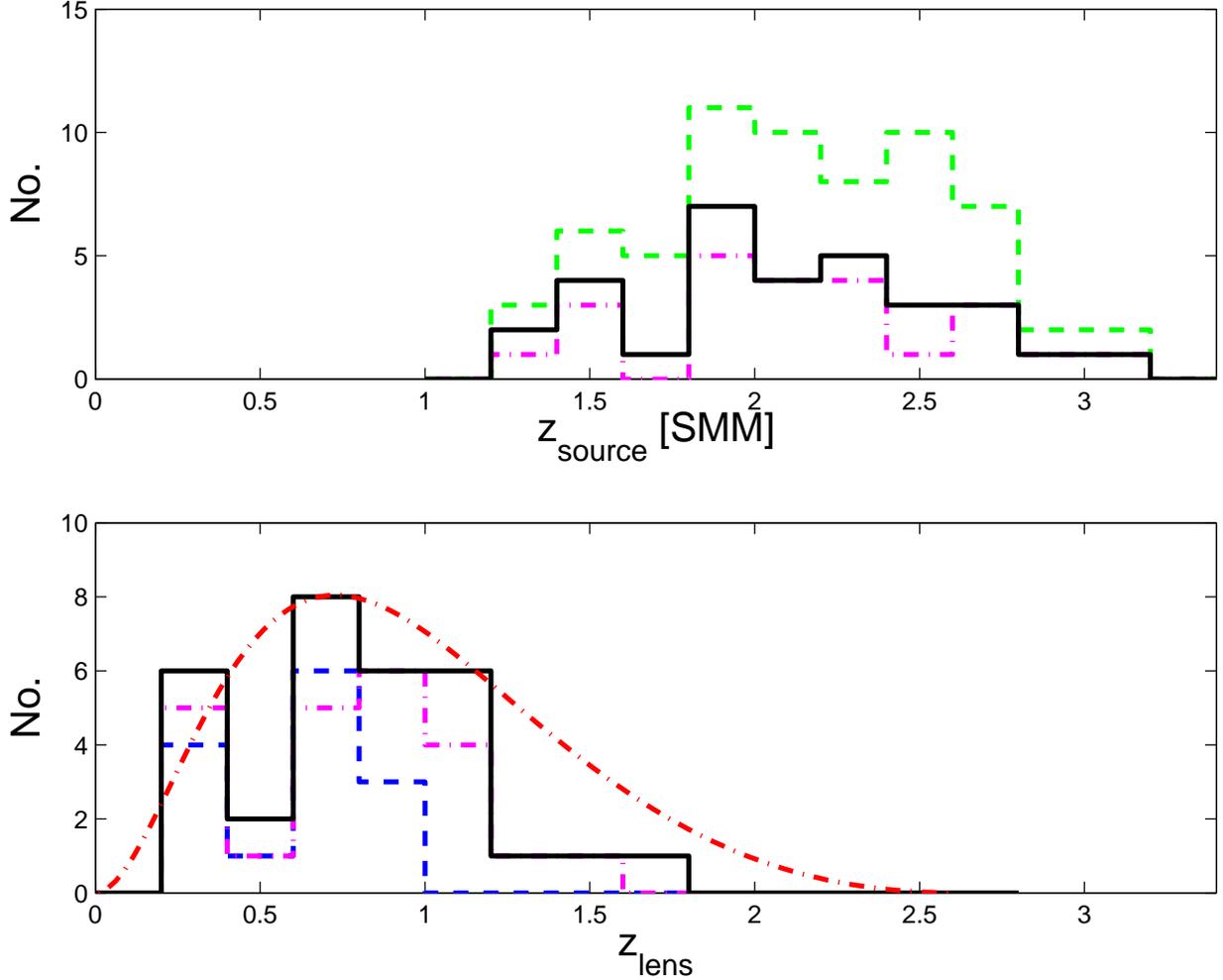}
\caption{Photometric redshift distribution of the SLG candidates (top panel) and of their VIKING associations (bottom panel) with (black solid histograms and without (magenta dot-dashed histogram) the 8 objects with estimated lens probability $<50\%$. In the upper panel, the green dashed line shows, for comparison, the redshift distribution for the full parent sample. In the lower panel, the blue dashed line shows the redshift distribution obtained using only the redshift estimates from the ANNz code (blue dashed line) while the red dot-dashed line shows the theoretical prediction from an updated version of the \citet{Negrello07} model for a source at $z=2.5$.}
\label{fig:zopt}%
\end{center}
\end{figure}

\begin{figure}
\begin{center}
\includegraphics[width=1\textwidth]{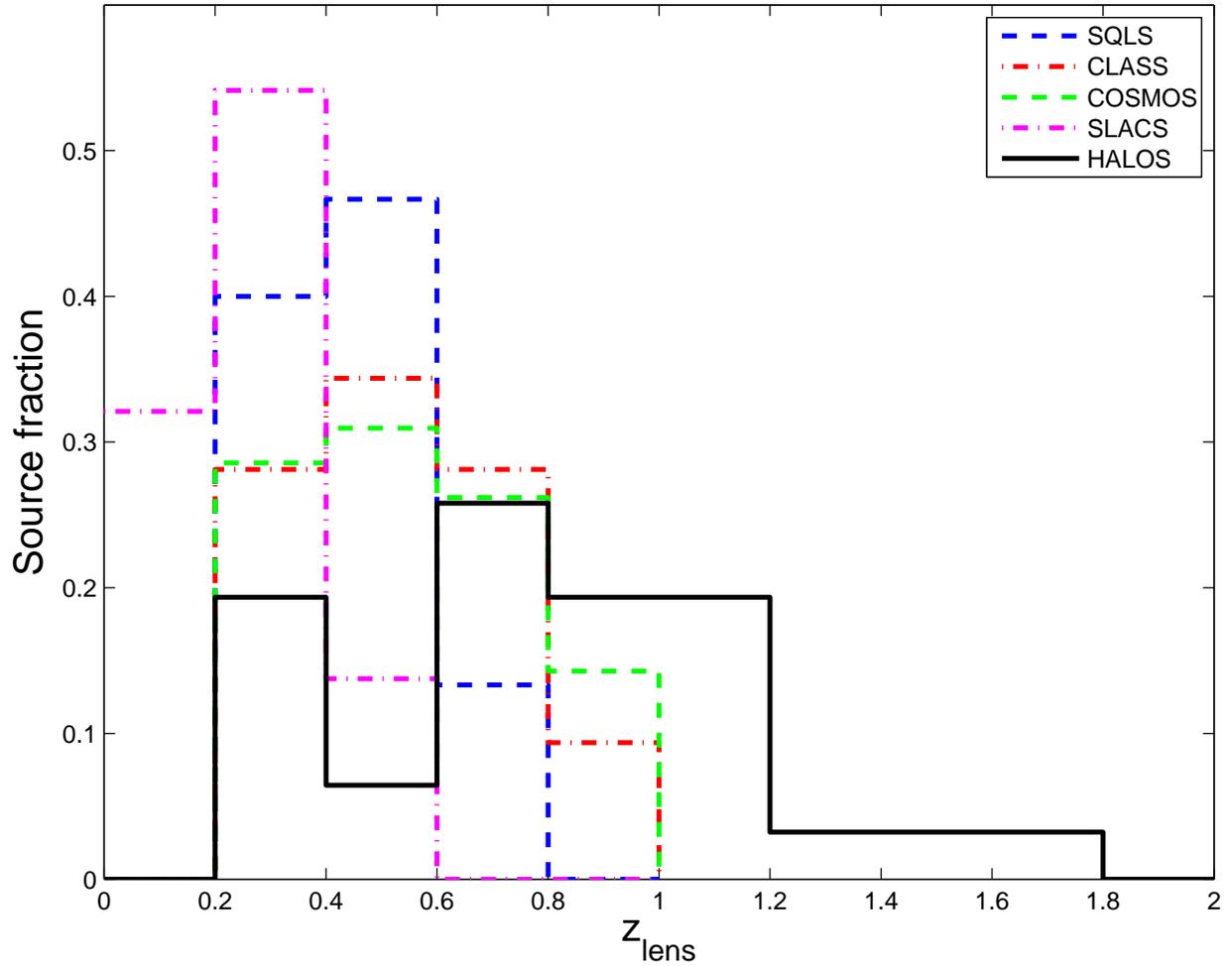}
\caption{Estimated normalized redshift distribution of lenses for our SLG candidates (31 objects; black solid line) compared with those of the CLASS (22 SLGs), COSMOS (20), SLACS (85), and SQLS (28) surveys, as given in Fig.~7 of \citet{Treu10}.}
\label{fig:treu}%
\end{center}
\end{figure}

\begin{figure}
\begin{center}
\includegraphics[width=1\textwidth]{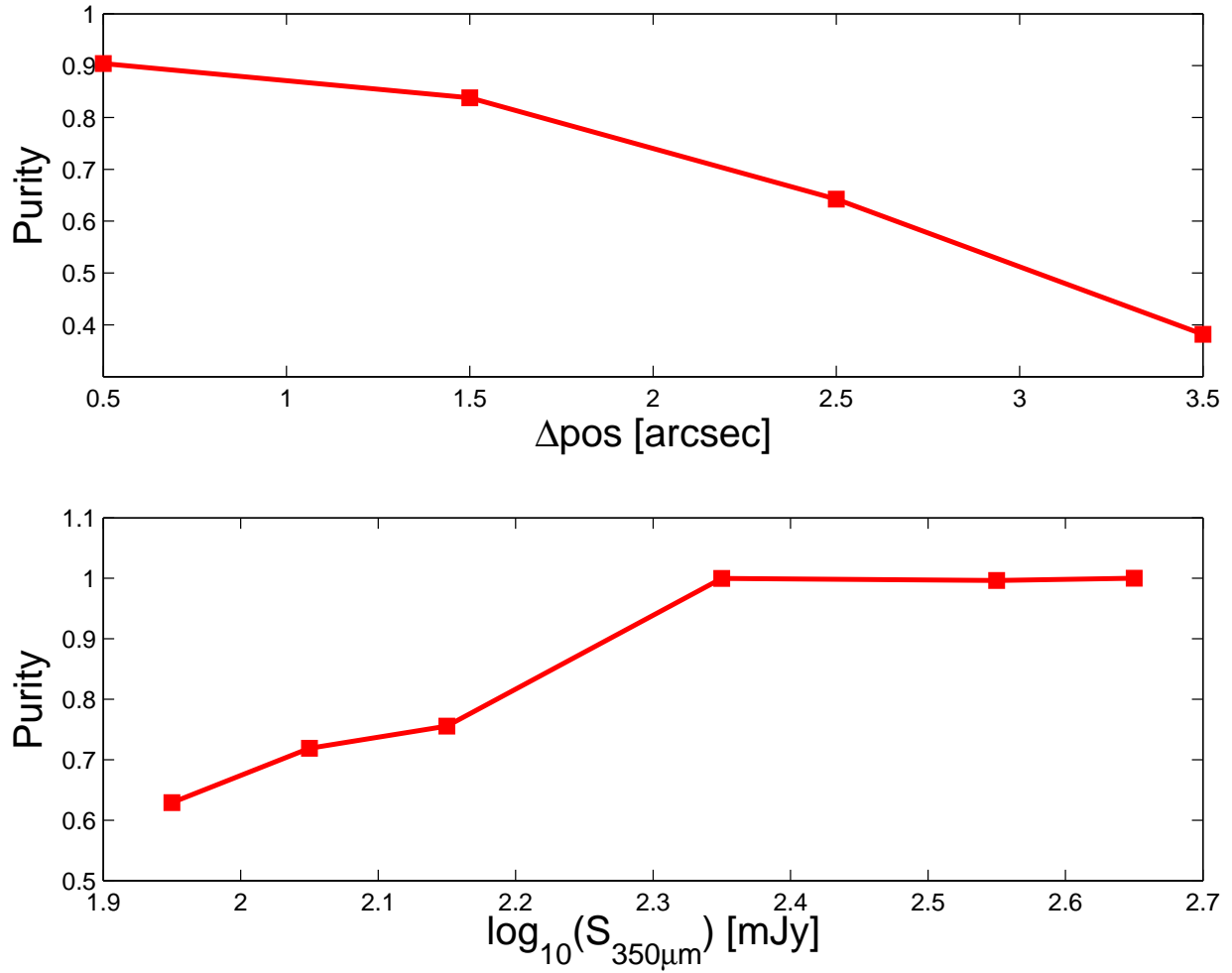}
\caption{Purity of the sample of candidate strongly lensed galaxies as a function of the angular separation between the candidate source and the candidate lens (upper panel) and of the $350\,\mu$m flux density of the candidate source (lower panel).}
\label{fig:purity}%
\end{center}
\end{figure}

\begin{figure}
\begin{center}
\includegraphics[width=1\textwidth]{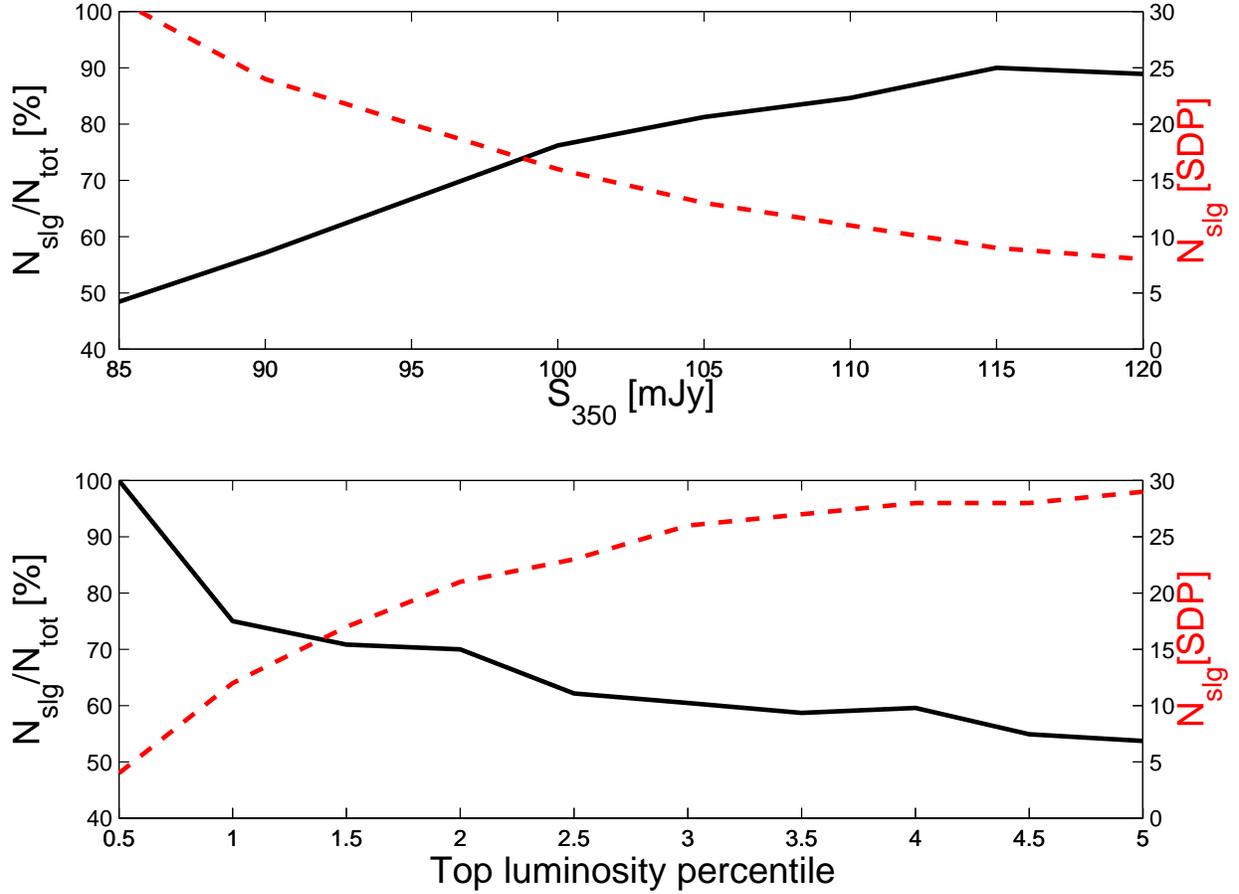}
\caption{Percentage (black solid line; left-hand scale) and total number (red dashed line; right-hand scale) of SLG candidates as a function of the $S_{350}$ threshold (top panel) and of the redshift-dependent apparent $100\,\mu$m luminosity threshold (bottom panel).
}
\label{fig:lumsel}%
\end{center}
\end{figure}

\begin{figure}
\begin{center}
\includegraphics[width=1\textwidth]{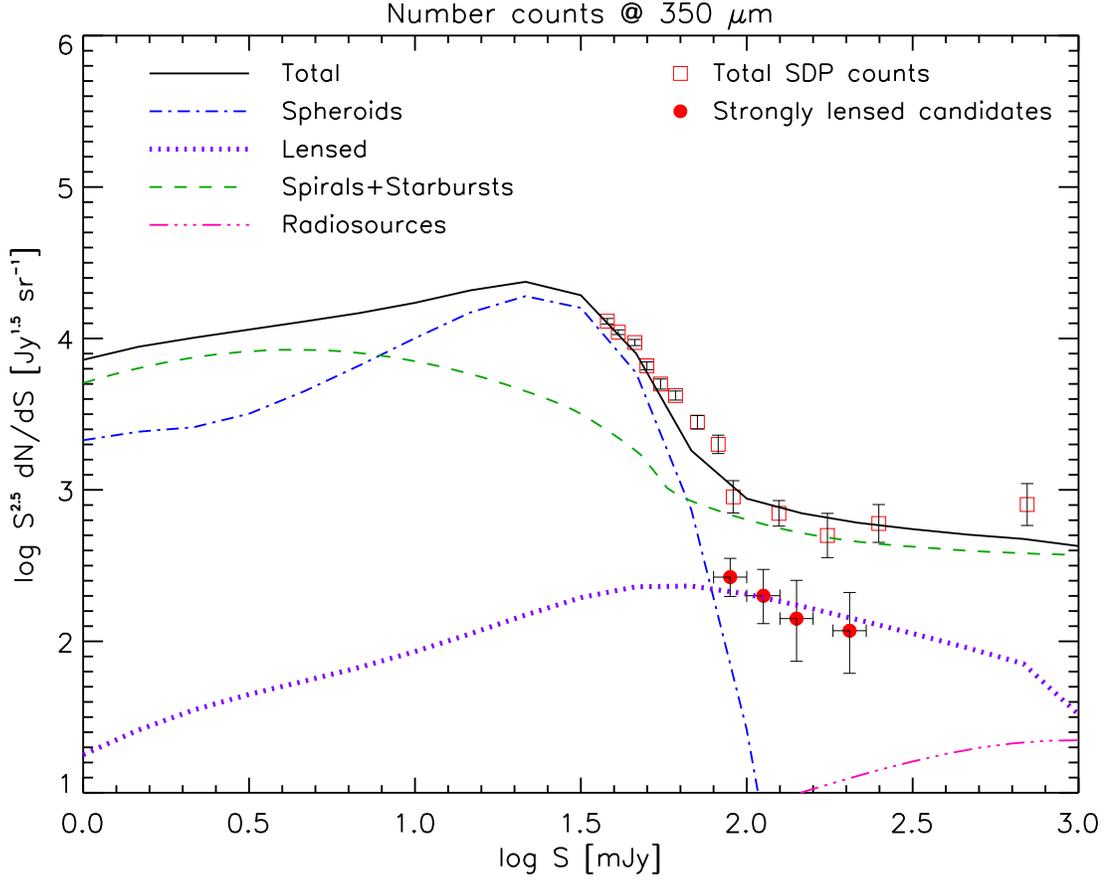}
\caption{Euclidean normalized differential number counts, corrected for the flux density dependent purity shown in Fig.~\protect\ref{fig:purity},  at $350\mu{\rm m}$ of the SLG candidates (filled red circles) compared with the total counts by \citet[][ open squares]{Clements10}. The errors are purely Poissonian. The lines show the contributions to the counts of lensed and unlensed star-forming galaxies as predicted by an updated lensing model by  \citet{Negrello07} coupled with the \citet{Lapi11} model for the evolution of the luminosity function, and of radio sources as yielded by the \citet{DeZotti05} model.}
\label{fig:SNC}%
\end{center}
\end{figure}

A preliminary estimate of the ``purity'' of the sample, using the available information on candidate sources and on candidate lenses, yield a global purity of 72\%. The estimated ``purity'' is found to decrease with increasing angular separation between the candidate source and the candidate lens and with decreasing flux density of candidate sources. The objects that will eventually turn out not to be SLGs would be in any case interesting targets for follow-up: Fig.~\ref{fig:LF} shows that all our SLG candidates have apparent star-formation rates of thousands ${\rm M}_\odot\,\hbox{yr}^{-1}$, i.e. apparent far-IR luminosities of a few to several times $10^{13}\,L_\odot$. Therefore if their emission is not amplified by a factor $\ge 2$, they would be among the brightest ultra-luminous infrared galaxies. 

\acknowledgments
We are grateful to the referee for a careful reading of the manuscript and useful comments. The work has been supported in part by ASI/INAF agreements I/009/10/0 (``Support for data analysis") and I/072/09/0 (``\textit{Planck} LFI Activity of Phase E2''), by INAF through the PRIN 2009 ``New light on the early Universe with sub-mm spectroscopy'', and by MIUR through the PRIN 2009. JGN acknowledge partial financial support from the Spanish Ministerio de Ciencia e Innovacion project AYA2010-21766-C03-01.

\bibliographystyle{apj}

\bibliography{strlen_finder}


\clearpage




\clearpage

\clearpage

\clearpage

\clearpage

\clearpage

\clearpage

\clearpage

\clearpage

\clearpage

\clearpage

\clearpage

\clearpage

\clearpage
\begin{center}

\begin{deluxetable}{lrccccccccc} \label{tab:slg_list}
\rotate
\tabletypesize{\scriptsize}
\tablewidth{0pt}
\tablecaption{The parent sample (\S\,\ref{sec:sample}). The $100\mu$m luminosity, $L_{100\mu{\rm m}}$, is in W/Hz. The 31 SLG candidates (see \S\,\ref{sec:optcrit}) are shown in boldface. Errors in parentheses.\label{tab:slg_list}}
\tablehead{
 \colhead{H-ATLAS} & \colhead{SDP} & \multicolumn{3}{c}{S[mJy]} & \colhead{$z_{\rm source}$} & \colhead{$\log(L_{100\mu{\rm m}}$)} & \colhead{$z_{\rm lens}$} & \colhead{($g-r$)} & \colhead{($Z-H$)} & \colhead{$\Delta$pos} \\
 & \colhead{ID} & \colhead{500$\mu$m} & \colhead{350$\mu$m} & \colhead{250$\mu$m} & \colhead{[SMM]} & \colhead{[SMM]} & & \colhead{[SDSS]} & \colhead{[VIKING]} & \colhead{[arcsec]}}
\startdata
\textbf{J090740.0-004200} & \textbf{  9\tablenotemark{N}} & \textbf{183 ( 9)} & \textbf{358 ( 8)} & \textbf{507 ( 7)} & \textbf{1.58 (0.01)} & \textbf{27.38 (0.30)} & \textbf{0.69 (0.13)$_{V }$} & \textbf{ 1.33} & \textbf{ 1.11} & \textbf{ 0.34} \\
\textbf{J091043.0-000321} & \textbf{ 11\tablenotemark{N}} & \textbf{249 (10)} & \textbf{403 ( 8)} & \textbf{462 ( 7)} & \textbf{1.79 (0.01)} & \textbf{27.59 (0.31)} & \textbf{0.72 (0.37)$_{V }$} & \textbf{ 1.93} & \textbf{ 1.23} & \textbf{ 0.98} \\
\textbf{J090302.9-014127} & \textbf{ 17\tablenotemark{N}} & \textbf{230 ( 9)} & \textbf{342 ( 8)} & \textbf{343 ( 7)} & \textbf{3.04 (0.01)} & \textbf{27.60 (0.28)} & \textbf{0.68 (0.40)$_{V }$} & \textbf{ 0.72} & \textbf{ 1.31} & \textbf{ 1.77} \\
\textbf{J091331.3-003642} & \textbf{ 44} & \textbf{ 89 (10)} & \textbf{151 ( 8)} & \textbf{187 ( 7)} & \textbf{1.50 (0.30)} & \textbf{27.03 (0.30)} & \textbf{0.30 (0.01)$_{S }$} & \textbf{ 1.45} & \textbf{ 0.88} & \textbf{ 1.79} \\
\textbf{J090051.0+015049} & \textbf{ 53} & \textbf{ 54 (10)} & \textbf{120 ( 8)} & \textbf{177 ( 7)} & \textbf{1.32 (0.28)} & \textbf{26.92 (0.32)} & \textbf{1.01 (0.09)$_{H }$} & \textbf{  --} & \textbf{ 1.79} & \textbf{ 1.40} \\
\textbf{J090952.9-010811} & \textbf{ 60} & \textbf{ 90 ( 9)} & \textbf{131 ( 8)} & \textbf{159 ( 7)} & \textbf{1.54 (0.40)} & \textbf{26.99 (0.38)} & \textbf{0.22 (0.02)$_{V }$} & \textbf{ 1.09} & \textbf{ 0.94} & \textbf{ 3.03} \\
J091341.4-004342 &  62 &  72 ( 9) & 124 ( 8) & 159 ( 7) & 1.53 (0.30) & 26.98 (0.28) & 1.07 (0.09)$_{H }$ &   -- &  1.53 &  3.85 \\
\textbf{J090957.6-003619} & \textbf{ 72\tablenotemark{a,b}} & \textbf{ 83 ( 9)} & \textbf{129 ( 8)} & \textbf{132 ( 7)} & \textbf{1.91 (0.46)} & \textbf{27.09 (0.39)} & \textbf{0.47 (0.18)$_{V }$} & \textbf{ 0.91} & \textbf{ 0.70} & \textbf{ 1.91} \\
\textbf{J090749.7-003807} & \textbf{ 79} & \textbf{ 69 ( 9)} & \textbf{113 ( 8)} & \textbf{140 ( 7)} & \textbf{1.48 (0.34)} & \textbf{26.90 (0.34)} & \textbf{1.05 (0.08)$_{H }$} & \textbf{  --} & \textbf{ 1.41} & \textbf{ 2.50} \\
\textbf{J090311.6+003906} & \textbf{ 81\tablenotemark{N}} & \textbf{173 (10)} & \textbf{202 ( 8)} & \textbf{135 ( 7)} & \textbf{2.63 (0.01)} & \textbf{27.60 (0.31)} & \textbf{0.30 (0.01)$_{S }$} & \textbf{ 1.67} & \textbf{ 0.82} & \textbf{ 0.73} \\
J090356.8+002310 &  87 &  64 ( 9) & 115 ( 8) & 131 ( 7) & 1.42 (0.36) & 26.84 (0.37) & 0.42 (0.05)$_{V }$ &  2.22 &  1.71 &  4.21 \\
\textbf{J090448.8+021646} & \textbf{ 98\tablenotemark{b}} & \textbf{ 52 ( 9)} & \textbf{ 97 ( 8)} & \textbf{126 ( 7)} & \textbf{1.47 (0.29)} & \textbf{26.85 (0.29)} & \textbf{0.63 (0.07)$_{H }$} & \textbf{  --} & \textbf{  --} & \textbf{ 3.10} \\
J090033.8+001957 & 103 &  44 ( 9) &  94 ( 8) & 125 ( 7) & 1.33 (0.28) & 26.77 (0.31) & 0.55 (0.16)$_{V }$ &  1.58 &  1.37 &  5.97 \\
J090459.3+020837 & 104 &  66 ( 9) & 102 ( 8) & 111 ( 7) & 1.79 (0.41) & 26.96 (0.39) & 0.34 (0.06)$_{V }$ &  0.53 &  0.04 &  9.43 \\
\textbf{J091056.5-002919} & \textbf{122} & \textbf{ 54 ( 9)} & \textbf{ 98 ( 8)} & \textbf{116 ( 7)} & \textbf{1.68 (0.31)} & \textbf{26.93 (0.29)} & \textbf{1.04 (0.14)$_{H }$} & \textbf{  --} & \textbf{ 1.95} & \textbf{ 2.35} \\
\textbf{J091521.5-002443} & \textbf{126} & \textbf{ 82 ( 9)} & \textbf{106 ( 8)} & \textbf{116 ( 7)} & \textbf{2.19 (0.42)} & \textbf{27.15 (0.31)} & \textbf{1.01 (0.36)$_{H }$} & \textbf{  --} & \textbf{  --} & \textbf{ 0.98} \\
\textbf{J090542.0+020733} & \textbf{127} & \textbf{ 58 (10)} & \textbf{100 ( 8)} & \textbf{112 ( 7)} & \textbf{1.82 (0.33)} & \textbf{26.98 (0.31)} & \textbf{0.87 (0.32)$_{V }$} & \textbf{ 0.39} & \textbf{  --} & \textbf{ 1.02} \\
\textbf{J091304.9-005343} & \textbf{130\tablenotemark{N}} & \textbf{112 ( 9)} & \textbf{141 ( 8)} & \textbf{110 ( 7)} & \textbf{2.30 (0.01)} & \textbf{27.39 (0.29)} & \textbf{0.26 (0.01)$_{S }$} & \textbf{ 1.28} & \textbf{ 0.80} & \textbf{ 2.35} \\
\textbf{J090626.6+022612} & \textbf{132} & \textbf{ 71 ( 9)} & \textbf{ 99 ( 8)} & \textbf{113 ( 7)} & \textbf{2.01 (0.42)} & \textbf{27.06 (0.34)} & \textbf{0.84 (0.14)$_{H }$} & \textbf{ 0.36} & \textbf{ 0.97} & \textbf{ 1.11} \\
J090459.9+015043 & 153 &  62 ( 9) &  88 ( 8) & 110 ( 7) & 1.74 (0.42) & 26.93 (0.40) & 0.36 (0.03)$_{V }$ &  1.27 &  1.05 &  6.57 \\
J090408.6+012610 & 180 &  46 (10) &  95 ( 8) & 103 ( 7) & 1.79 (0.36) & 26.93 (0.34) & 1.29 (0.20)$_{H }$ &   -- &   -- &  2.51 \\
J090403.9+005619 & 183 &  50 ( 9) &  87 ( 8) & 104 ( 7) & 1.69 (0.31) & 26.88 (0.30) & 0.31 (0.02)$_{V }$ &  1.04 &  0.88 &  4.08 \\
J090653.3+023207 & 189 &  64 ( 9) &  95 ( 8) & 102 ( 7) & 2.08 (0.37) & 27.05 (0.29) & 1.02 (0.10)$_{H }$ &  1.25 &  1.88 &  5.16 \\
\textbf{J090707.9-003134} & \textbf{191} & \textbf{ 56 (10)} & \textbf{ 95 ( 8)} & \textbf{102 ( 7)} & \textbf{1.94 (0.34)} & \textbf{26.99 (0.30)} & \textbf{1.11 (0.06)$_{H }$} & \textbf{  --} & \textbf{  --} & \textbf{ 2.01} \\
J091305.1-001409 & 194 &  62 ( 9) &  97 ( 8) &  95 ( 7) & 2.18 (0.37) & 27.06 (0.27) & 0.47 (0.04)$_{V }$ &  2.50 &  1.06 &  7.31 \\
\textbf{J090732.3-005207} & \textbf{217} & \textbf{ 52 ( 9)} & \textbf{103 ( 8)} & \textbf{ 91 ( 7)} & \textbf{2.13 (0.47)} & \textbf{27.02 (0.35)} & \textbf{1.13 (0.16)$_{H }$} & \textbf{  --} & \textbf{ 1.62} & \textbf{ 2.78} \\
J091354.6-004539 & 219 &  50 ( 9) &  87 ( 8) &  92 ( 7) & 1.95 (0.34) & 26.95 (0.30) & 0.42 (0.10)$_{V }$ &  1.50 &  0.72 &  5.67 \\
J090504.8+000800 & 225 &  71 ( 9) &  97 ( 8) &  97 ( 7) & 2.18 (0.46) & 27.07 (0.33) &  --  &   -- &   -- &   -- \\
J090308.3-000420 & 227 &  55 (10) &  95 ( 8) &  99 ( 7) & 1.98 (0.35) & 26.99 (0.29) & 0.88 (0.07)$_{H }$ &   -- &  1.44 &  7.47 \\
\textbf{J090705.7+002128} & \textbf{237} & \textbf{ 64 (10)} & \textbf{ 92 ( 8)} & \textbf{ 90 ( 7)} & \textbf{2.28 (0.38)} & \textbf{27.08 (0.27)} & \textbf{1.37 (0.23)$_{H }$} & \textbf{  --} & \textbf{  --} & \textbf{ 2.56} \\
J090433.4-010740 & 238 &  70 ( 9) & 104 ( 8) &  84 ( 7) & 2.52 (0.44) & 27.17 (0.33) & 0.15 (0.12)$_{VH}$ &  0.61 &  0.19 &  7.58 \\
J090239.0+002819 & 249 &  48 ( 9) &  85 ( 8) &  97 ( 7) & 1.81 (0.32) & 26.91 (0.31) & 0.43 (0.16)$_{H }$ &   -- &  1.04 &  8.03 \\
\textbf{J090931.8+000133} & \textbf{257\tablenotemark{a,b}} & \textbf{ 56 (10)} & \textbf{ 88 ( 8)} & \textbf{ 92 ( 7)} & \textbf{1.91 (0.54)} & \textbf{26.93 (0.45)} & \textbf{0.34 (0.21)$_{V }$} & \textbf{ 0.46} & \textbf{ 0.51} & \textbf{ 2.37} \\
\textbf{J090459.0-012911} & \textbf{265} & \textbf{ 44 ( 9)} & \textbf{ 88 ( 8)} & \textbf{ 92 ( 7)} & \textbf{1.87 (0.37)} & \textbf{26.91 (0.34)} & \textbf{0.97 (0.07)$_{H }$} & \textbf{  --} & \textbf{ 1.32} & \textbf{ 2.30} \\
\textbf{J091148.2+003355} & \textbf{290\tablenotemark{b}} & \textbf{ 72 ( 9)} & \textbf{102 ( 8)} & \textbf{ 90 ( 7)} & \textbf{2.45 (0.40)} & \textbf{27.16 (0.29)} & \textbf{0.74 (0.20)$_{V }$} & \textbf{ 1.41} & \textbf{  --} & \textbf{ 3.05} \\
\textbf{J090319.6+015635} & \textbf{301} & \textbf{ 61 ( 9)} & \textbf{ 89 ( 8)} & \textbf{ 87 ( 7)} & \textbf{2.26 (0.38)} & \textbf{27.06 (0.27)} & \textbf{0.82 (0.07)$_{H }$} & \textbf{  --} & \textbf{ 1.30} & \textbf{ 2.20} \\
J090405.3-003331 & 302 &  76 (10) &  98 ( 8) &  85 ( 7) & 2.57 (0.41) & 27.19 (0.31) & 0.66 (0.08)$_{VH}$ &  0.67 &  1.04 &  4.42 \\
\textbf{J085751.3+013334} & \textbf{309} & \textbf{ 64 (10)} & \textbf{ 90 ( 8)} & \textbf{ 88 ( 7)} & \textbf{2.29 (0.38)} & \textbf{27.08 (0.28)} & \textbf{1.78 (0.10)$_{H }$} & \textbf{  --} & \textbf{  --} & \textbf{ 2.11} \\
J085900.3+001405 & 312 &  50 (10) &  92 ( 8) &  89 ( 7) & 2.06 (0.39) & 26.98 (0.31) &  --  &   -- &   -- &   -- \\
\textbf{J091351.7-002340} & \textbf{327} & \textbf{ 48 (10)} & \textbf{ 89 ( 8)} & \textbf{ 90 ( 7)} & \textbf{1.99 (0.37)} & \textbf{26.95 (0.31)} & \textbf{0.88 (0.42)$_{V }$} & \textbf{ 0.47} & \textbf{ 1.19} & \textbf{ 1.89} \\
J090446.4+022218 & 329 &  61 ( 9) &  86 ( 8) &  87 ( 7) & 2.24 (0.38) & 27.05 (0.28) & 0.26 (0.05)$_{V }$ &  0.67 &  0.67 &  9.09 \\
J091003.5+021028 & 340 &  46 ( 9) &  86 ( 8) &  87 ( 7) & 1.97 (0.37) & 26.93 (0.31) &  --  &   -- &   -- &   -- \\
J090429.6+002935 & 354\tablenotemark{c} &  60 (10) &  86 ( 8) &  85 ( 7) & 2.18 (0.60) & 27.01 (0.44) & 0.18 (0.01)$_{V }$ &  3.32 &  0.62 &  3.16 \\
J090032.7+004316 & 383 &  50 (10) &  85 ( 8) &  85 ( 7) & 2.08 (0.36) & 26.97 (0.28) & 0.52 (0.14)$_{V }$ &  0.98 &  1.24 &  3.85 \\
J090613.7-010044 & 390 &  69 ( 9) &  86 ( 8) &  79 ( 7) & 2.52 (0.41) & 27.14 (0.31) &  --  &   -- &   -- &   -- \\
\textbf{J090453.2+022018} & \textbf{392} & \textbf{ 88 ( 9)} & \textbf{107 ( 8)} & \textbf{ 83 ( 7)} & \textbf{2.80 (0.44)} & \textbf{27.28 (0.29)} & \textbf{0.66 (0.11)$_{V }$} & \textbf{ 2.01} & \textbf{ 1.02} & \textbf{ 1.97} \\
\textbf{J085855.3+013728} & \textbf{393} & \textbf{ 68 ( 9)} & \textbf{ 92 ( 8)} & \textbf{ 80 ( 7)} & \textbf{2.51 (0.40)} & \textbf{27.14 (0.30)} & \textbf{1.48 (0.17)$_{H }$} & \textbf{  --} & \textbf{ 2.35} & \textbf{ 1.43} \\
J090346.1+013428 & 396 &  68 ( 9) &  91 ( 8) &  79 ( 7) & 2.53 (0.41) & 27.15 (0.30) &  --  &   -- &   -- &   -- \\
\textbf{J090954.6+001754} & \textbf{407} & \textbf{ 76 (10)} & \textbf{111 ( 8)} & \textbf{ 83 ( 7)} & \textbf{2.67 (0.48)} & \textbf{27.23 (0.33)} & \textbf{0.68 (0.25)$_{V }$} & \textbf{  --} & \textbf{ 1.09} & \textbf{ 2.11} \\
J090440.0-013439 & 414 &  76 ( 9) &  98 ( 8) &  76 ( 7) & 2.75 (0.44) & 27.22 (0.30) &  --  &   -- &   -- &   -- \\
\textbf{J090950.8+000427} & \textbf{419\tablenotemark{a}} & \textbf{ 68 ( 9)} & \textbf{ 92 ( 8)} & \textbf{ 81 ( 7)} & \textbf{2.50 (0.40)} & \textbf{27.14 (0.30)} & \textbf{0.62 (0.31)$_{V }$} & \textbf{ 0.69} & \textbf{ 0.37} & \textbf{ 2.46} \\
J090204.1-003829 & 436 &  61 ( 9) &  87 ( 8) &  79 ( 7) & 2.40 (0.39) & 27.08 (0.29) &  --  &   -- &   -- &   -- \\
J090930.4+002224 & 462 &  57 ( 9) &  85 ( 8) &  76 ( 7) & 2.39 (0.40) & 27.06 (0.29) & 1.00 (0.10)$_{H }$ &   -- &  1.71 &  4.89 \\
\textbf{J090409.4+010734} & \textbf{476} & \textbf{ 55 (10)} & \textbf{ 86 ( 8)} & \textbf{ 79 ( 7)} & \textbf{2.29 (0.39)} & \textbf{27.03 (0.28)} & \textbf{0.92 (0.09)$_{VH}$} & \textbf{ 0.22} & \textbf{ 1.54} & \textbf{ 2.76} \\
J090310.6+015149 & 503 &  71 ( 9) & 105 ( 8) &  77 ( 7) & 2.69 (0.49) & 27.20 (0.34) & 1.00 (0.07)$_{H }$ &   -- &  1.66 &  4.26 \\
\textbf{J085859.2+002818} & \textbf{512} & \textbf{ 52 ( 9)} & \textbf{ 85 ( 8)} & \textbf{ 77 ( 7)} & \textbf{2.28 (0.41)} & \textbf{27.01 (0.29)} & \textbf{0.43 (0.03)$_{V }$} & \textbf{ 1.56} & \textbf{ 1.04} & \textbf{ 2.96} \\
J090530.4+012800 & 514 &  65 ( 9) &  90 ( 8) &  78 ( 7) & 2.51 (0.41) & 27.13 (0.30) & 0.49 (0.07)$_{V }$ &  2.84 &  1.01 &  5.38 \\
J090818.9+023330 & 515 &  57 (10) &  92 ( 8) &  77 ( 7) & 2.42 (0.44) & 27.08 (0.32) &  --  &   -- &   -- &   -- \\
J090441.5+015154 & 545\tablenotemark{c} &  70 (10) &  87 ( 8) &  76 ( 7) & 2.60 (0.42) & 27.16 (0.31) & -0.02 (0.02)$_{H }$ &   -- &  0.29 &  3.21 \\
\textbf{J090739.1-003948} & \textbf{639\tablenotemark{a}} & \textbf{ 81 ( 9)} & \textbf{ 99 ( 8)} & \textbf{ 73 ( 7)} & \textbf{2.89 (0.45)} & \textbf{27.25 (0.28)} & \textbf{0.39 (0.15)$_{H }$} & \textbf{  --} & \textbf{ 0.64} & \textbf{ 0.82} \\
J091257.2+000300 & 700 &  87 (10) &  96 ( 8) &  69 ( 7) & 3.03 (0.46) & 27.29 (0.27) &  --  &   -- &   -- &   -- \\
J090819.1-002026 & 751 &  66 (10) &  93 ( 8) &  69 ( 7) & 2.71 (0.47) & 27.16 (0.32) &  --  &   -- &   -- &   -- \\
J090813.0-003657 & 775 &  65 (10) &  88 ( 8) &  66 ( 7) & 2.74 (0.45) & 27.16 (0.31) &  --  &   -- &   -- &   -- \\
J085908.5+011320 & 910 &  70 (10) &  87 ( 8) &  67 ( 7) & 2.81 (0.44) & 27.19 (0.29) & 0.94 (0.08)$_{H }$ &   -- &  1.50 &  5.47 \\
\enddata
\tablenotetext{a}{The optical counterpart has colors compatible with those of a late-type galaxy.}
\tablenotetext{b}{Tentative lens probability $<30\%$.}
\tablenotetext{c}{The optical counterpart is local, $z<0.2$.}
\tablenotetext{N}{Confirmed strongly lensed galaxy \citep{Negrello10}.}
\tablecomments{The subscript next to the lens redshift indicates its origin: `V' for VIKING, `H' for H-ATLAS (this work), `VH' for galaxies whose redshifts were re-estimated by us because the VIKING estimate has exceedingly large errors ($z=1.83\pm 1.25$, $1.86\pm 1.42$, $1.56\pm 1.35$ for SDP.238, SDP.302, and SDP.476, respectively), and `S' for spectroscopic measurements.}
\end{deluxetable}
\end{center}




\end{document}